# New Mira Variables from the MACHO Galactic Bulge Fields, part II

KLAUS BERNHARD [1,3,4], STEFAN HÜMMERICH [2,3,4]

1) A-4030 Linz, Austria; e-mail: klaus . bernhard@liwest . at
2) D-56338 Braubach, Germany; e-mail: ernham@rz-online . de
3) Bundesdeutsche Arbeitsgemeinschaft für Veränderliche Sterne e.V. (BAV), Munsterdamm 90, D-12169 Berlin, Germany
4) American Association of Variable Star Observers (AAVSO), 49 Bay State Road, Cambridge, MA 02138, USA

BAV Mitteilungen Nr. 229

**Abstract:** We present a new sample of 525 Mira variables in the direction of the Galactic Bulge, expanding on previous samples of 69 and 500 objects, respectively, and thereby concluding our search for new Mira variables in the MACHO Galactic Bulge fields. 364 Miras of the present sample are reported as variable stars for the first time. We have cross-correlated our sample with the sample of Mira stars from the OGLE-III Catalog of Long-Period Variables (LPVs) in the Galactic Bulge and found 146 matches; MACHO and OGLE periods are in very good overall agreement. We present summary data for all stars of the present sample and give a statistical overview, comparing the properties of the MACHO and OGLE samples and enlarging on the analyses in our previous paper. Lightcurves, folded lightcurves and further details are available via the AAVSO International Variable Star Index (http://www.aavso.org/vsx/). Data of the complete sample of Mira variables from the MACHO Galactic Bulge fields, as presented in our papers (Bernhard, 2011; Huemmerich and Bernhard, 2012 and the present paper), can be found in the appendix.

1. Introduction

We have continued our search for Mira variables in the MACHO Galactic Bulge fields (Bernhard, 2011; Huemmerich and Bernhard, 2012). The MACHO project (http://macho.anu.edu.au/) comprises observations carried out between 1992 and 2000 with the 1.27 m Great Melbourne Telescope situated at Mount Stromlo in Australia. All observations were taken simultaneously through the non-standard MACHO blue filter (~4500-6300 Å; hereafter MACHO B-band) and MACHO red filter (~6300-7600 Å; hereafter MACHO R-band) using a combination of eight 2048*2048 CCD cameras (Alcock et al., 1999). For more information on the MACHO project see e.g. Alcock et al. (1997).

Retaining the methodology outlined in our first paper, we have inspected MACHO R-band lightcurves from the MACHO Galactic Bulge fields in order to find suitable Mira candidates. MACHO R-band was chosen over B-band photometry because of its increased sensitivity towards the red band of the electromagnetic spectrum, making it more suitable for identifying red variables such as Miras. In the case of three stars, however, it was necessary to fall back on B-band observations because of bad R-band photometry. We have then transformed MACHO instrumental magnitudes on to the Kron-Cousins system by using equation (2) of Alcock et al. (1999). Only stars with an amplitude > 2 mag (Rc) were investigated and subjected to a visual inspection of their lightcurves; objects exhibiting significant changes in amplitude, mean magnitude and / or period suggesting semi-regularity have been rejected. For the very bright objects, ASAS-3 V data (Pojmanski, 2002) has been included into the analysis whenever available in order to increase the time baseline and achieve a period solution of higher accuracy.

We have cross-matched our sample with the 2MASS Catalog (Skrutskie et al., 2006), from which we derive astrometric positions and near-infrared color indices. Each object was checked against the Strasbourg CDS Vizier service (Ochsenbein et al., 2000) and the AAVSO International Variable Star Index (Watson et al., 2006) for pre-existence as a Mira-type star in variability catalogs. In addition, we have established a cross-correlation with the sample of Mira stars from the OGLE-III Catalog of Long-Period Variables (LPVs) in the Galactic Bulge (Soszyński et al., 2013; hereafter OGLE sample), the results of which are presented in Chapter 2 along with a comparison of the MACHO and OGLE samples.

Summary data for all new Mira variables are presented in Table 1, which also gives corresponding identifiers from other lists. Lightcurves, folded lightcurves and further details are available via the AAVSO International Variable Star Index (Watson et al., 2006; http://www.aavso.org/vsx/). Data of the Mira variables from our previous papers can be found online at VizieR (catalog J/other/OEJV/149) and in the Peremennye Zvezdy Variable Stars Supplement (PZP, vol. 11, N 12). Additionally, data of the complete sample of Mira stars from the MACHO Galactic Bulge fields, as presented in our papers (Bernhard, 2011[1]; Huemmerich and Bernhard, 2012 and the present paper), can be found in the appendix, including 2MASS J, H, K photometry (Skrutskie et al., 2006).

## 2. Properties of the MACHO Mira sample and comparison with the OGLE sample

*2.1 Cross-correlation with Mira variables from the OGLE-III Catalog of Long-Period Variables (LPVs) in the Galactic Bulge*

We have cross-correlated the present sample of MACHO Miras with the OGLE sample (Soszyński et al., 2013). We find 146 matches, which is in agreement with our expectations as the sky coverage of the two surveys is different and many of the brighter MACHO objects will likely be saturated in the OGLE frames. Corresponding OGLE identifiers (OGLE-BLG-LPV-NNNNNN) are listed in Table 1.

Except for two cases, in which the given period was half the actual value[2], MACHO and OGLE periods are in very good overall agreement. In exactly 50% of cases, MACHO and OGLE periods agree to within 1%, while the period difference is more than 4.5% for only 5 stars of the entire sample (see Figure 1). Examples of excellent period agreement between MACHO and OGLE Miras are presented in Figure 2.

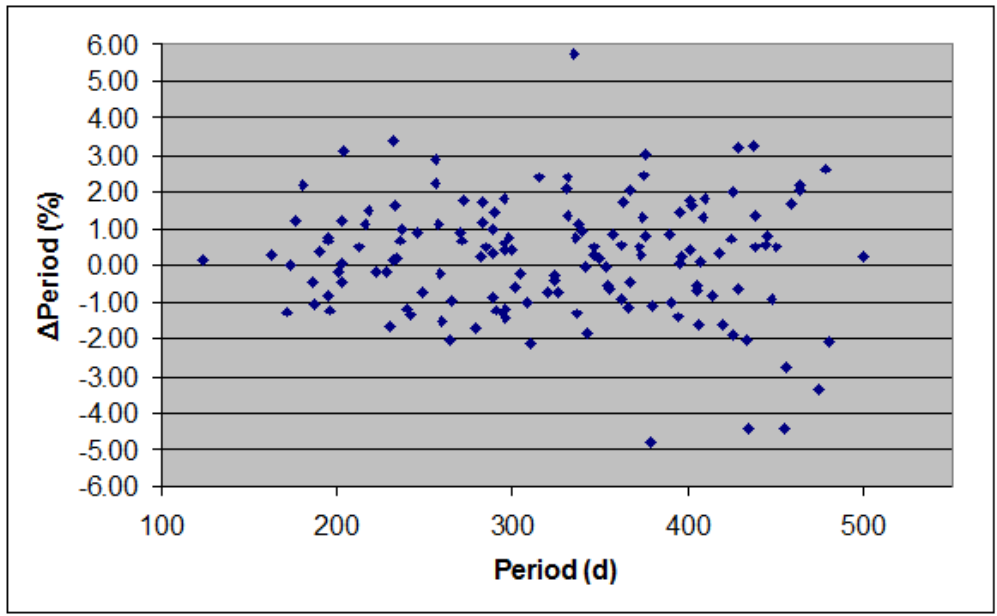

*Figure 1: Difference between OGLE and MACHO periods, as expressed in ΔPeriod (%)*

We have investigated all stars with ΔPeriod > 3% and find that the observed differences in period are mostly due to the disparity between MACHO and OGLE coverage. Both samples are based on quite heterogeneous datasets. OGLE time coverage and number of datapoints vary "[...] from about 100 points collected over two years to more than 3,000 observations obtained between 1997 and 2009 (OGLE-II + OGLE-III)." (Soszyński et al., 2013). MACHO data, on the other hand, comprises from about 200 to more than 1,200 datapoints which were mostly collected over a timespan from 1,500 to 2,500 days in the case of the Galactic Bulge fields. MACHO fields with fewer than 200 observations have been excluded from our analysis.

---

[1] One star of the sample presented in Bernhard (2011) has been identified as a duplicate entry (MACHO 101.20779.46 = MACHO 104.20779.6004). As better coverage of the object has been achieved in field 101, we retain MACHO 101.20779.46 and drop the other identifier from our sample.

[2] Both objects (MACHO 180.22111.49, P = 182.5 d in the MACHO sample; OGLE-BLG-LPV-211391, P = 155.96 d in the OGLE sample) are listed with their corrected periods (P = 365 d and P = 305 d, respectively) in Table 1.

As indicated above, longer time coverage results in more accurate results in almost all cases. This was expected, as small cycle-to-cycle variations, which Mira variables are notorious for, may have a great impact on the period solutions for stars with short time baselines that only cover a small number of cycles. This holds especially true for long-period Miras, for which analyses are sometimes based on only two consecutive maxima. Furthermore, in some cases, maxima have been covered only partially or not at all, which is seen frequently in stars whose periods are very nearly equal to one year. Additionally, varying lightcurve shapes add uncertainty to the period analysis. Considering these difficulties, and the fact that Mira variables are prone to exhibiting intrinsic period scatter (cf. e.g. Koen and Lombard, 1995; Zijlstra and Bedding, 2002), the excellent agreement of MACHO and OGLE periods is noteworthy. Figures 3 and 4 give examples of the period solutions for MACHO and OGLE Miras with ΔPeriod > 3% which illustrate the frequent disparity in time coverage between both datasets.

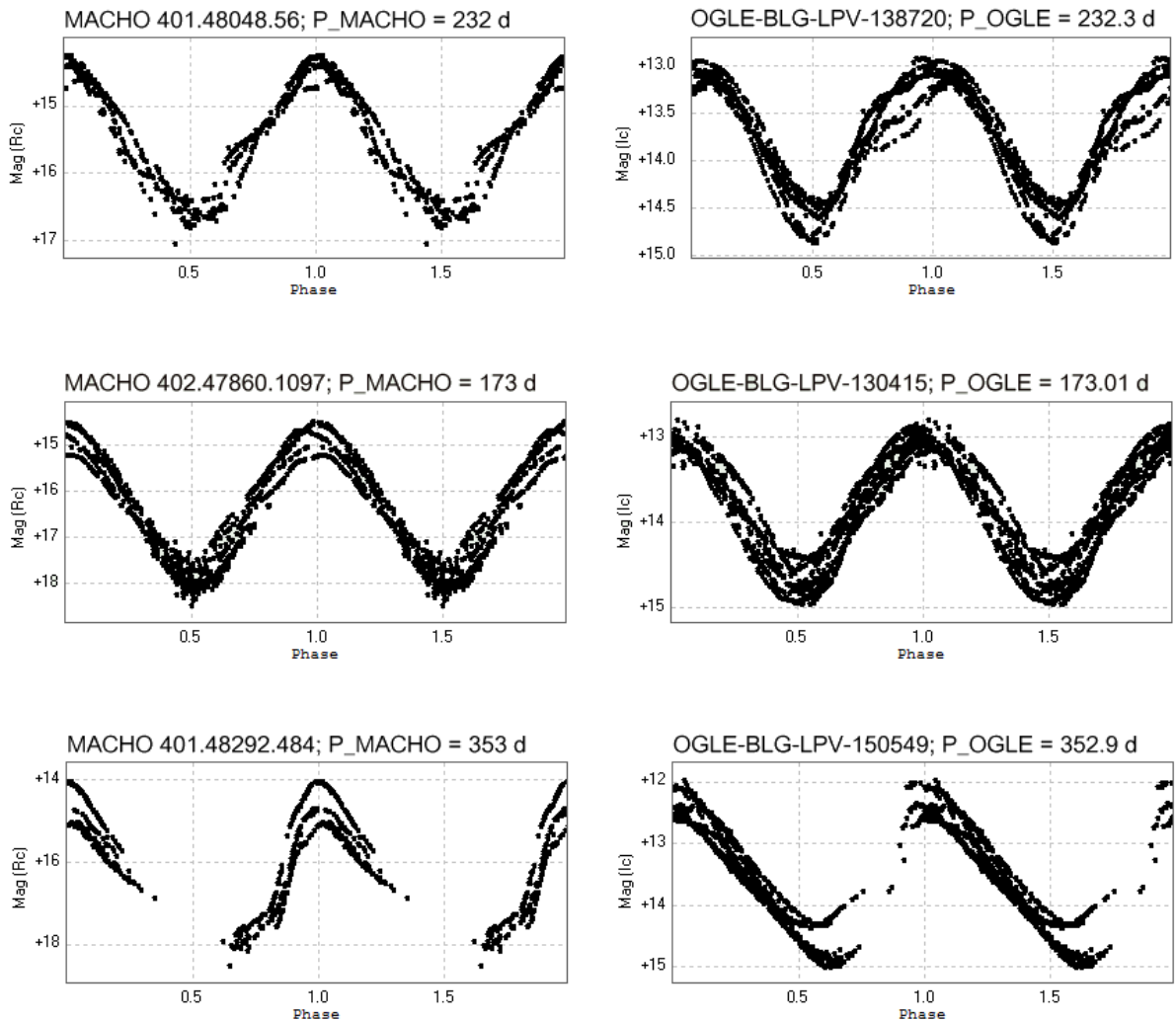

*Figure 2: Period solutions for the stars MACHO 401.48048.56 / OGLE-BLG-LPV-138720 (top), MACHO 402.47860.1097 / OGLE-BLG-LPV-130415 (middle), MACHO 401.48292.484 / OGLE-BLG-LPV-150549 (bottom), based on MACHO data (left side) and OGLE data (right side). Although MACHO and OGLE data were taken at different epochs, there is excellent agreement between MACHO and OGLE periods, which also indicates the stability of the pulsational behaviour of these Miras.*

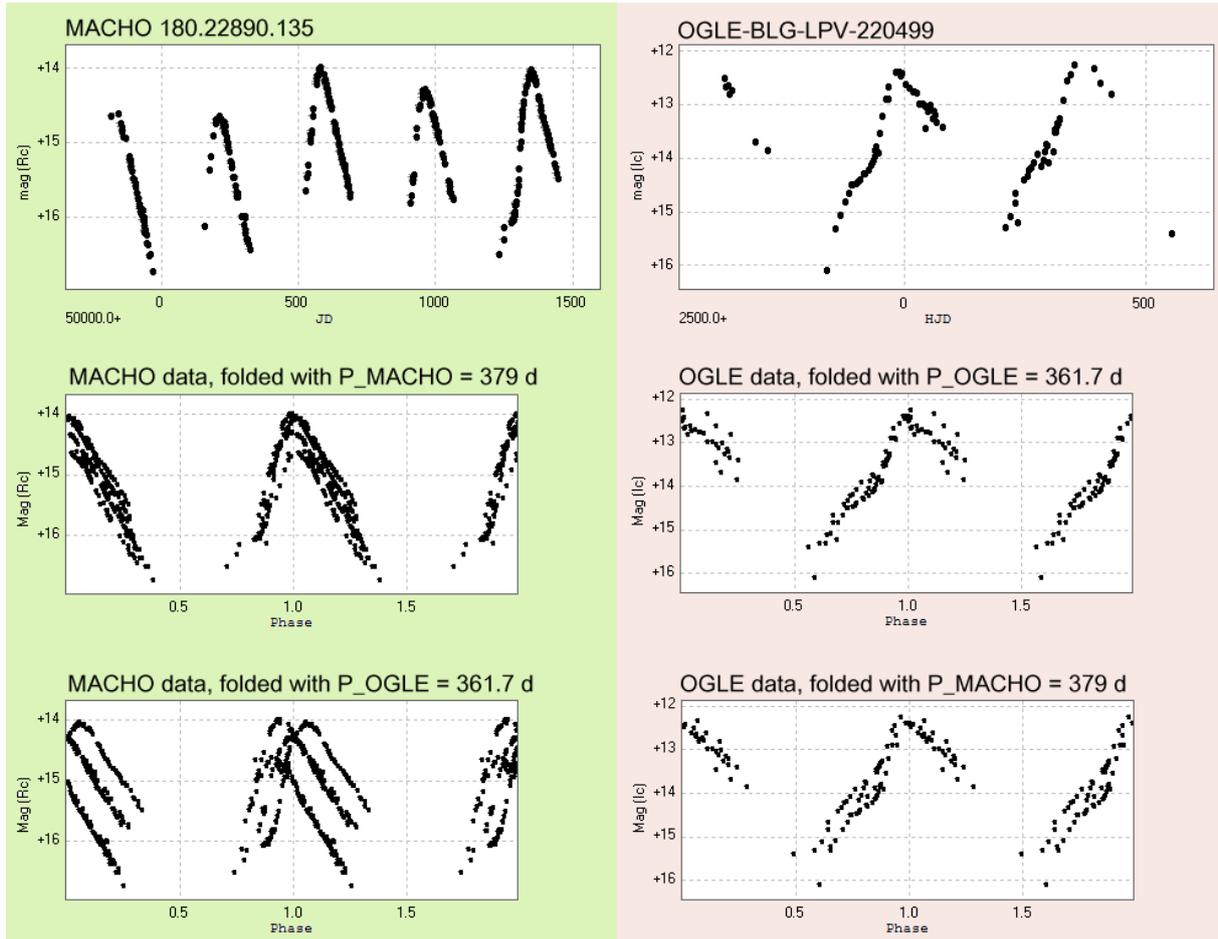

*Figure 3: The star MACHO 180.22890.135 / OGLE-BLG-LPV-220499. The left side of the figure shows MACHO Rc data (green-shaded area): lightcurve (top) and corresponding phase plots based on MACHO period (middle) and OGLE period (bottom). The right side of the figure shows OGLE Ic data (red-shaded area): lightcurve (top) and corresponding phase plots based on OGLE period (middle) and MACHO period (bottom). The MACHO period produces a better fit of both datasets, which is to be expected because of longer time coverage.*

In agreement with the above mentioned deductions, the disparity in coverage seems to be most pronounced for objects with ΔPeriod > 3%. We have therefore chosen to augment this situation by combining MACHO and OGLE data for all objects with ΔPeriod > 3%. The resulting increment of the time baseline enables us to obtain a period solution of higher accuracy for these stars; an example of this procedure is illustrated in Figure 5. Stars, whose periods result from a combination of MACHO and OGLE data, are marked as such in Table 1.

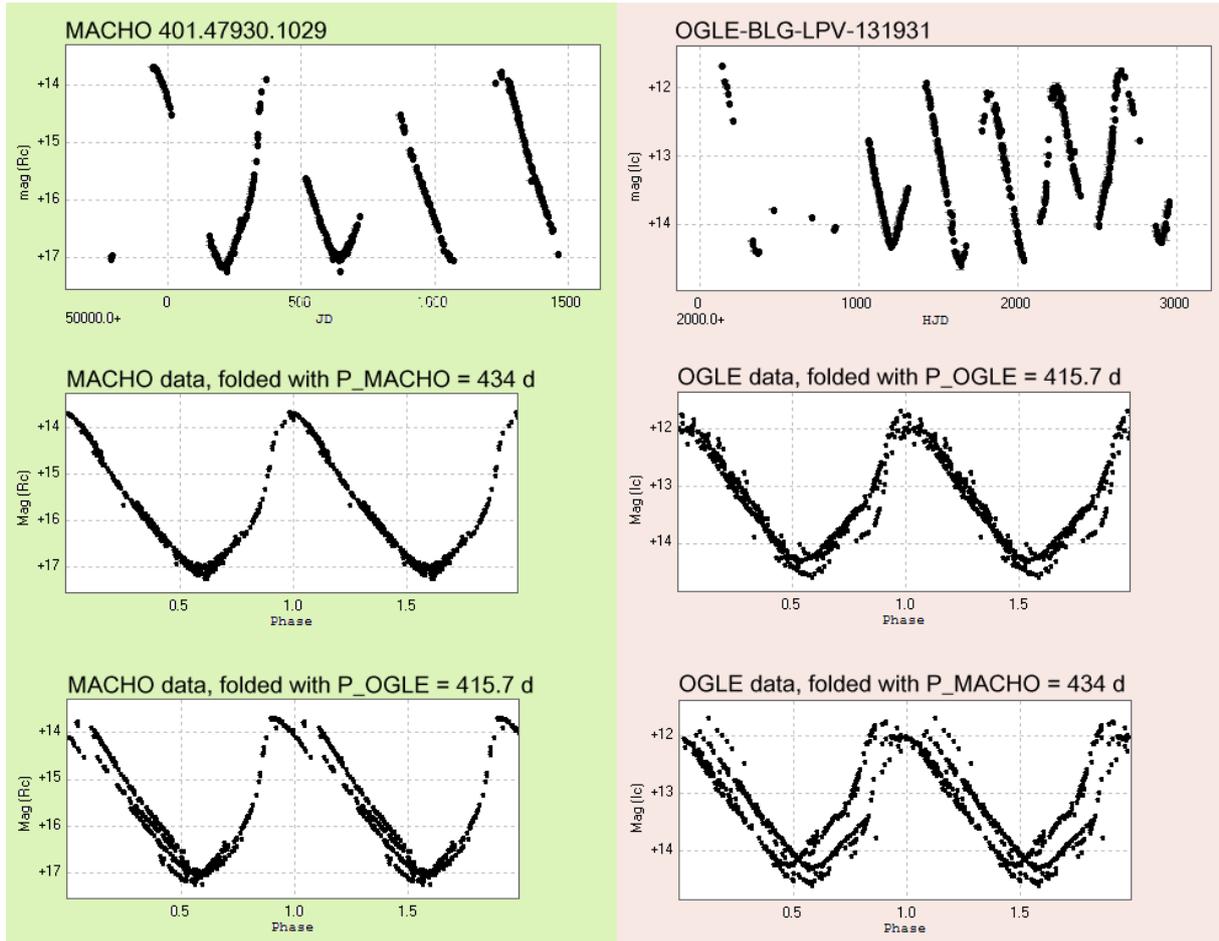

*Figure 4: The star MACHO 401.47930.1029 / OGLE-BLG-LPV-131931. The specification and arrangement of the data are the same as in Figure 3. Both period solutions produce a better fit of their respective data sets. The period solution of the OGLE sample is preferable in this case as it has been based on a considerably larger dataset which also boasts better coverage of maxima.*

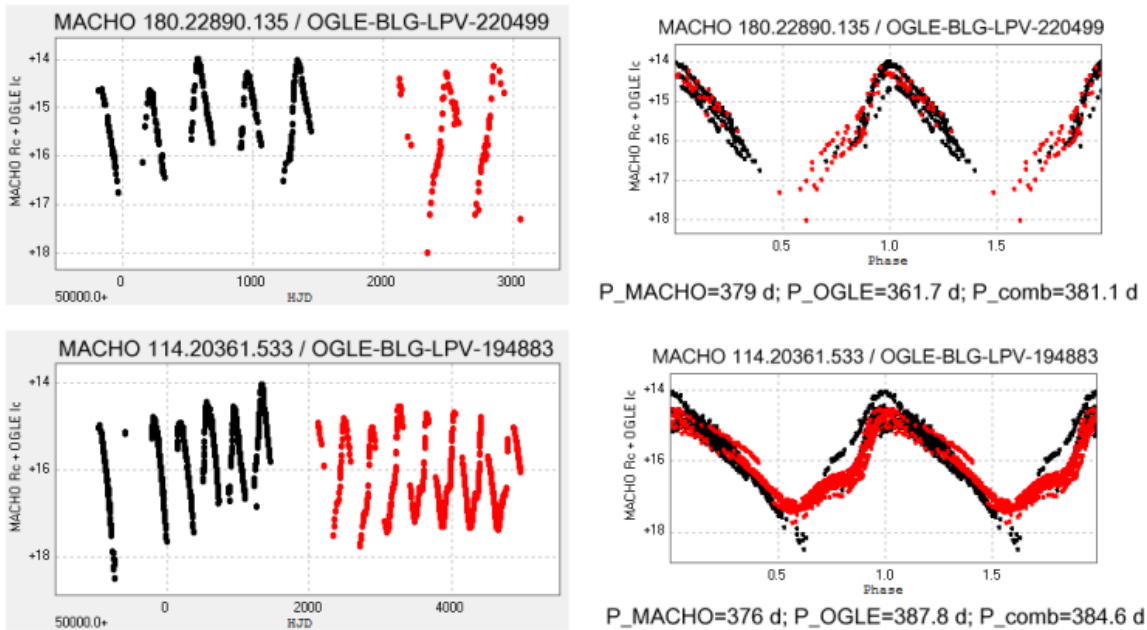

*Figure 5: Period solutions for the stars MACHO 180.22890.135 / OGLE-BLG-LPV-220499 (top) and MACHO 114.20361.533 / OGLE-BLG-LPV-194883 (bottom), based on a combination of MACHO and OGLE data. For the period analysis, OGLE Ic (red) has been shifted to match MACHO Rc (black).*

*2.2 Period distribution*[3]

We have compared the period distribution of the MACHO and OGLE samples, the result of which is illustrated in Figure 6. The OGLE sample (N = 6528) is more complete and comprises about six times as many Mira variables as the MACHO sample (N = 1094), which bears on the following results. Nevertheless, and despite of an overall good agreement, there is a noteworthy discrepancy in the period distribution between both samples. The MACHO sample contains more Miras with periods ranging from 201-350 days, notably in the range from 201-300 days. In contrast, the OGLE sample encompasses a much higher percentage of Miras in the period range > 350 days. In fact, the longest-period Mira we have been able to identify in MACHO data is the OH maser source MACHO 305.35072.100 with a period of P = 592 d (cf. also Huemmerich and Bernhard, 2012); there do not seem to be Miras of longer period in the entire MACHO sample.

The observed discrepancy is most likely due to the different passbands and limiting magnitudes of the MACHO and OGLE projects. OGLE observations are taken in the Cousins I-band (Ic), which roughly comprises a wavelength region between ~6800 and ~9000 Å (cf. e.g. Moro and Munari, 2000), and are thus much more suited to finding long-period Miras which are mostly very red objects due to extinction by circumstellar dust (cf. e.g. Matsunaga et al., 2005). Furthermore, a fraction of the brighter Miras will likely be saturated in the OGLE frames, which possibly contributes to the observed differences in period distribution between both samples.

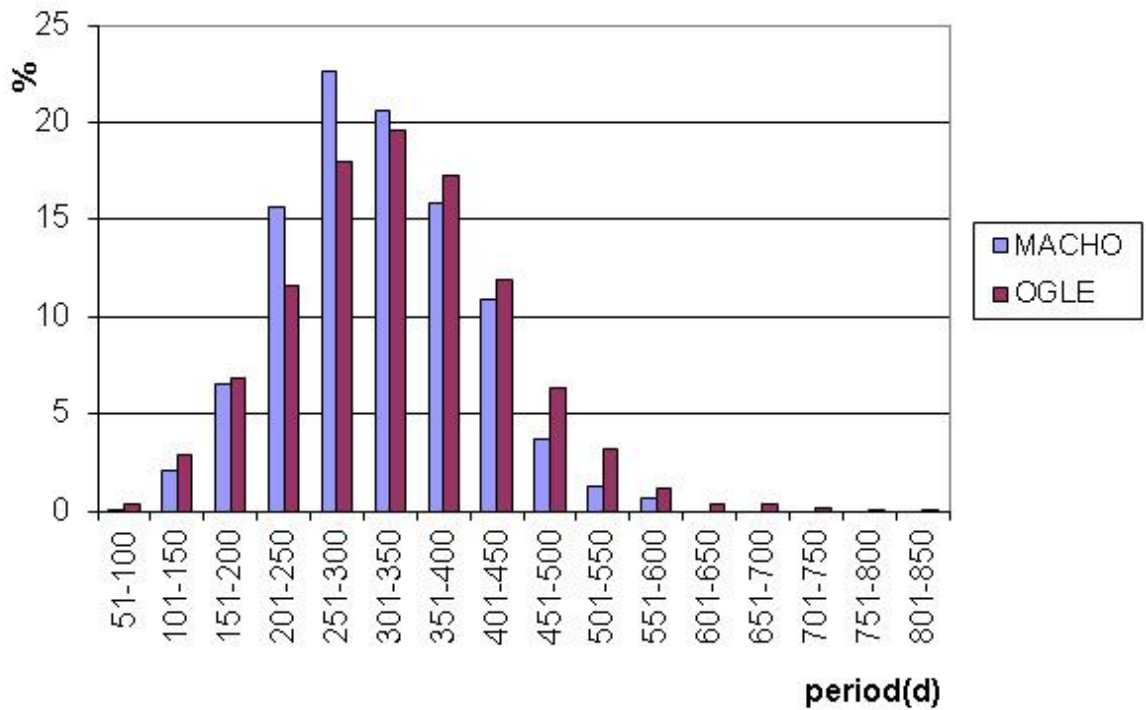

*Figure 6: Period distribution of Mira variables in the direction of the Galactic Bulge (lilac: MACHO sample; N = 1094 / purple: OGLE sample; N = 6528)*

---

[3] All analyses presented in this chapter and the following ones are based on the whole sample of Mira variables from the MACHO Galactic Bulge fields as presented in Bernhard (2011), Huemmerich and Bernhard (2012) and the present paper.

*2.3 Properties in colour-magnitude, period-colour and period-luminosity space*

*2.3.1 Colour-magnitude diagrams*

Figure 7 shows the colour-magnitude diagrams for Mira variables from the MACHO and OGLE samples, the results of which are in excellent agreement. Both diagrams exhibit a red tail of Miras with (H-Ks) ≥ 1, extending to about (H-Ks) = 1.5 in the case of MACHO Miras and (H-Ks) = 2 in the case of the OGLE sample. This is in agreement with the findings of Matsunaga et al. (2005); cf. in particular their Figure 7. It is noteworthy that Miras with (H-Ks) ≥ 1 become fainter with increasing (H-Ks), which becomes especially obvious in the OGLE sample, demonstrating again the advantages of OGLE in discovering Miras towards the red and faint end. This underluminosity in the 2MASS Ks-band is likely caused by circumstellar extinction due to dust (cf. e.g. Fraser, 2008).

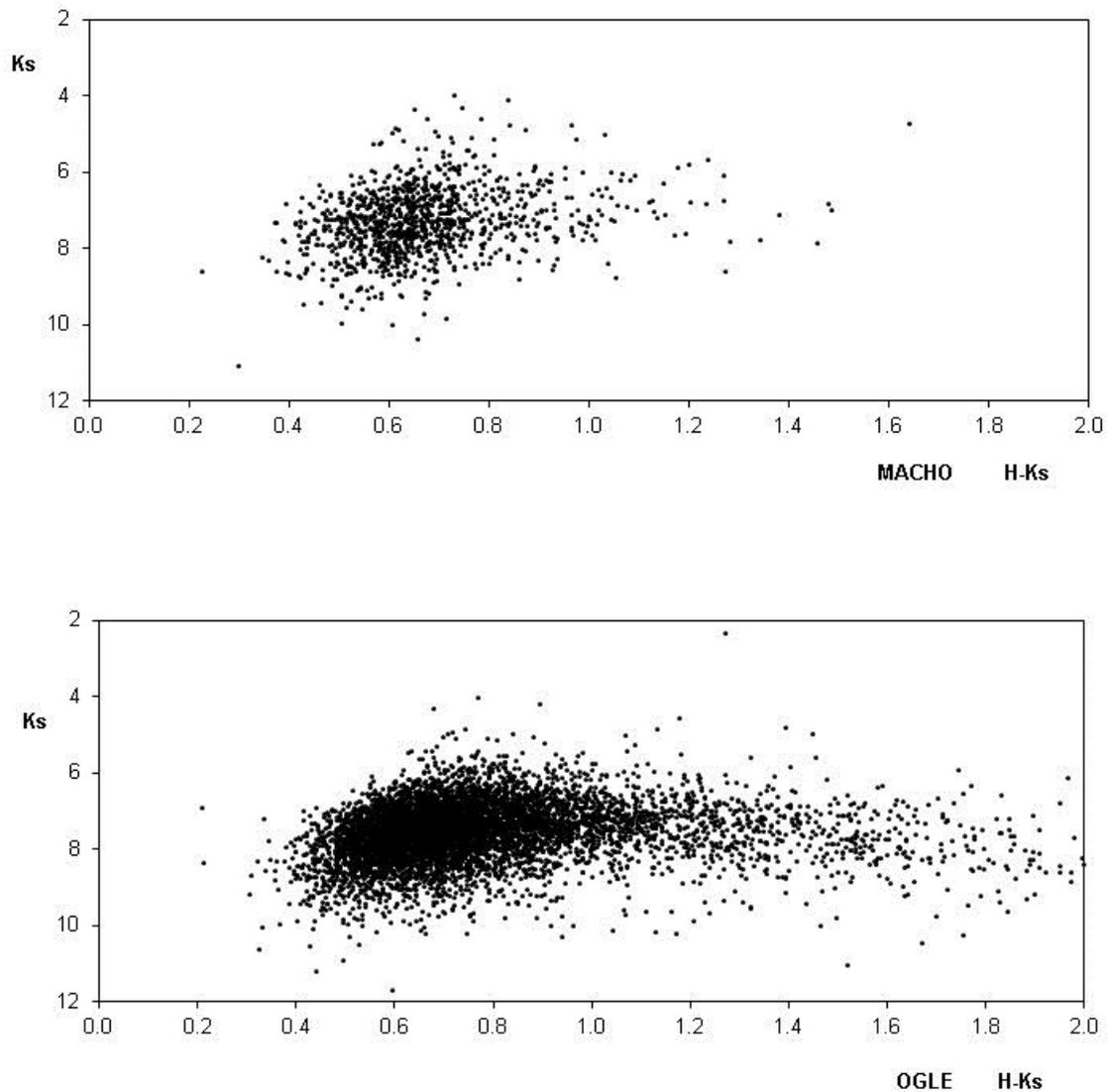

*Figure 7: 2MASS (H-Ks) vs. Ks diagrams for the MACHO sample (top; N = 1094) and the OGLE sample (bottom; N = 6528)*

*2.3.2 Period-colour diagrams*

Period-colour diagrams of the MACHO and OGLE samples are given in Figure 8; period is expressed in log (P). As expected, Miras of longer period have larger (H-Ks) values and thus redder colours. There is a turn-off point at log (P) ~ 2.6, at which the increase in (H-Ks) follows a steeper slope, indicating that Miras with periods longer than log (P) ~ 2.6 tend to show colour excesses because of circumstellar dust shells, as denoted by Matsunaga et al. (2005).

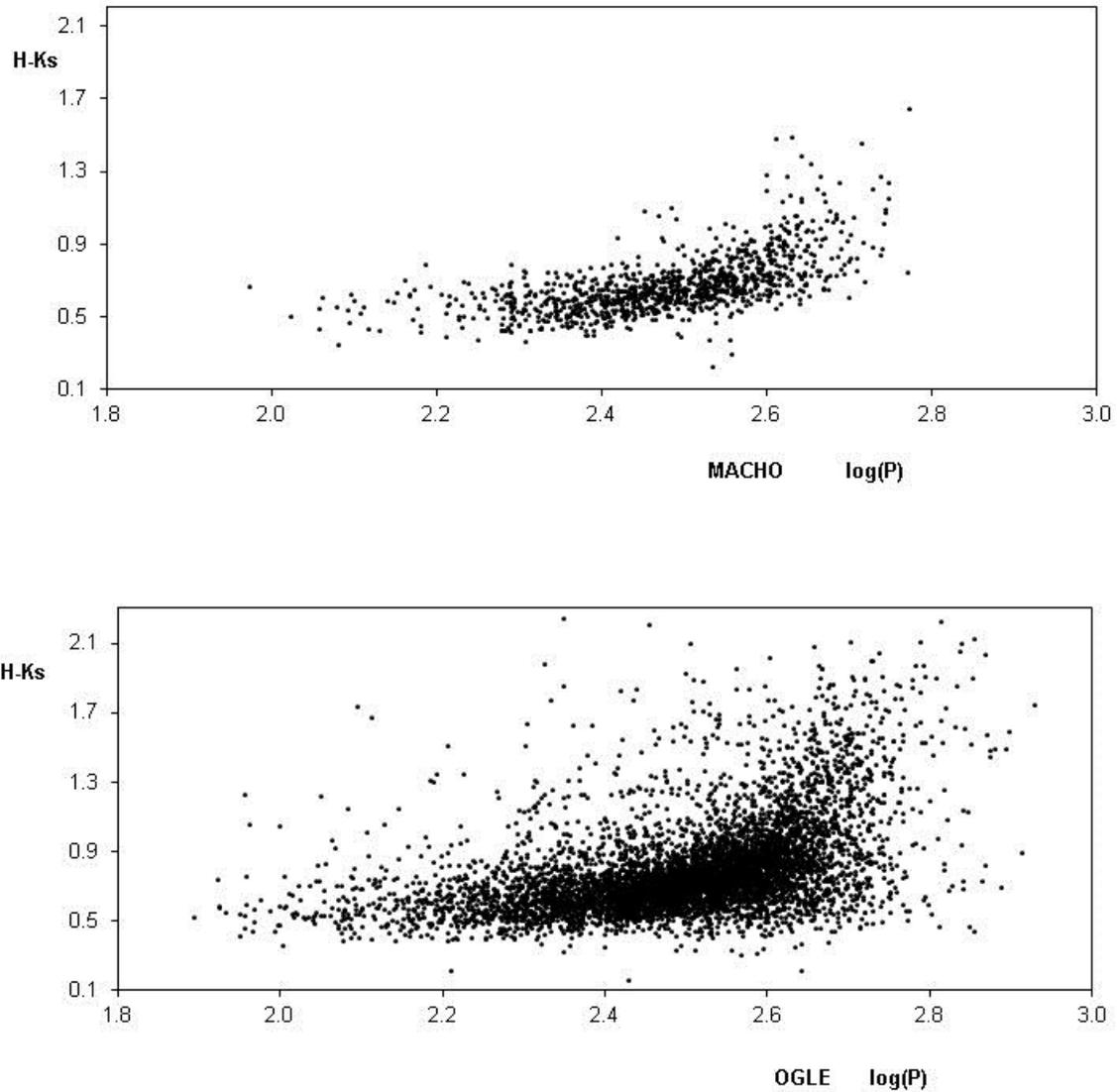

*Figure 8: Log (P) vs 2MASS (H-Ks) diagrams for the MACHO sample (top; N = 1094) and the OGLE sample (bottom; N = 6528)*

*2.3.3 Period-luminosity diagrams*

The distribution of Mira variables in the period-luminosity plane is rather clearly outlined. They occupy what has become to be known as "sequence C" and are well separated from the semi-regular and OSARG (OGLE Small Amplitude Red Giant) variables (cf. e.g. Wood et al., 1999; Soszyński et al., 2013; especially their Figure 5). There is excellent agreement in the distribution of Mira variables in the period-luminosity diagrams for both samples, with OGLE data reaching to fainter Ks.

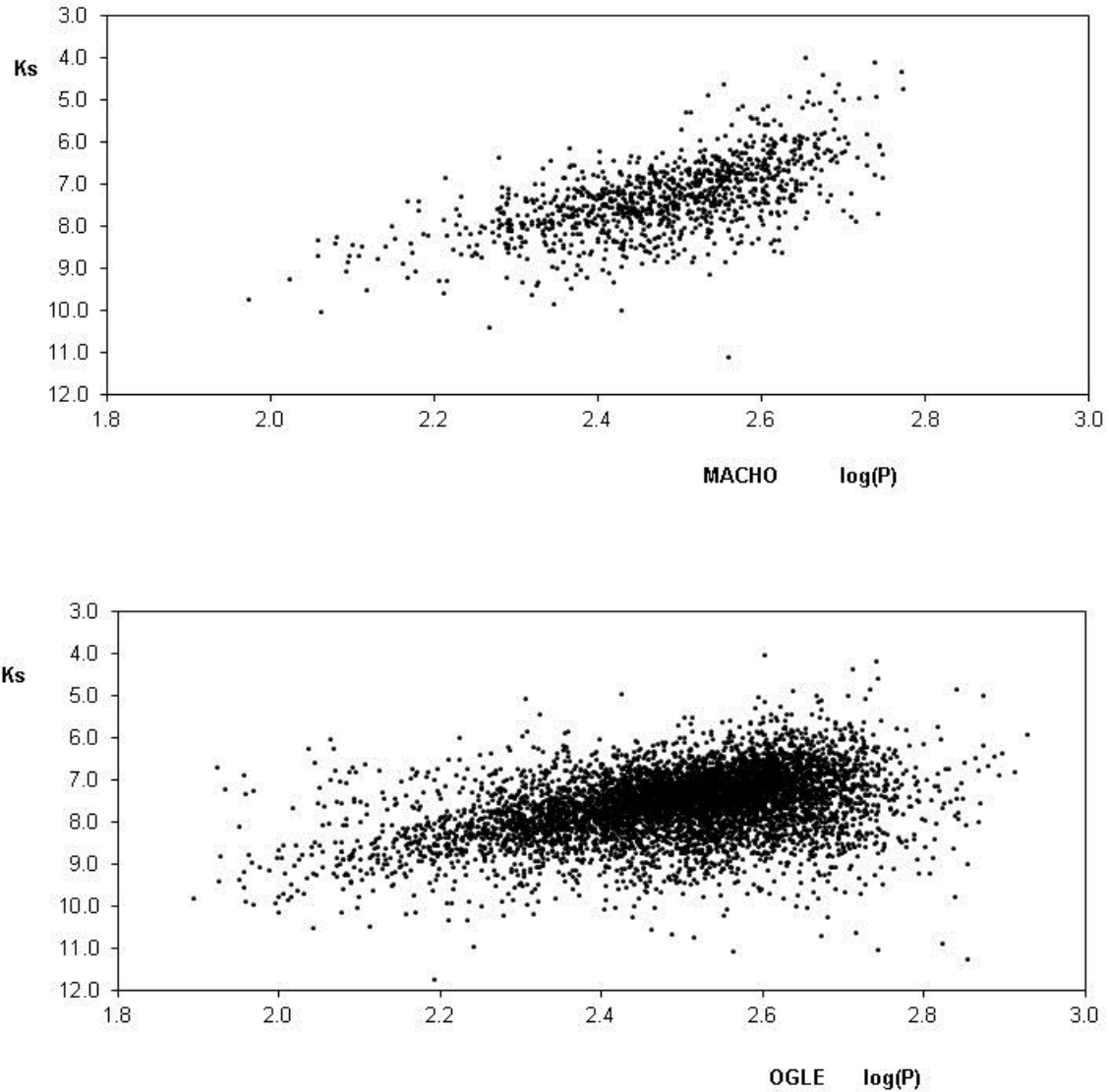

*Figure 9: Log (P) vs 2MASS Ks diagrams for the MACHO sample (top; N = 1094) and the OGLE sample (bottom; N = 6528)*

Table 1: Summary data for the new sample of 525 Miras from the MACHO Galactic Bulge fields

| No. | MACHO | RA (J2000) | DEC (J2000) | Range(MACHO) | | Epoch(Max) | Per(d) | Other ID / Remarks |
|---|---|---|---|---|---|---|---|---|
| 501 | 403.47552.12 | 17 54 44.10 | -29 27 50.1 | 12.3-15.3 | Rc | 2451254 | 331 | OGLE-BLG-LPV-114862 |
| 502 | 402.47618.372 | 17 55 17.01 | -29 04 50.8 | 13.1-17.1 | Rc | 2451268 | 197 | |
| 503 | 402.47624.8 | 17 55 17.77 | -28 40 43.5 | 12.2-<16.4 | Rc | 2450959 | 241 | |
| 504 | 402.47678.275 | 17 55 27.31 | -29 02 33.2 | 12.2-<16.7 | Rc | 2450885 | 268 | |
| 505 | 402.47680.779 | 17 55 27.40 | -28 57 43.8 | 14.1-17.8: | Rc | 2451055 | 400 | |
| 506 | 402.47675.8 | 17 55 30.05 | -29 15 10.0 | >12.4-<15.5 | Rc | 2451028 | 203 | OGLE-BLG-LPV-121843 |
| 507 | 402.47676.58 | 17 55 32.71 | -29 11 58.6 | >13.4-16.8 | Rc | 2451427 | 312 | |
| 508 | 402.47683.5 | 17 55 33.86 | -28 44 55.7 | 12.0-15.7 | Rc | 2450632 | 197 | |
| 509 | 402.47684.48 | 17 55 35.08 | -28 41 15.1 | >11.4:-17.0 | Rc | 2451405 | 547 | |
| 510 | 402.47683.10 | 17 55 35.92 | -28 43 01.9 | 12.7-16.5 | Rc | 2451279 | 215 | |
| 511 | 402.47739.56 | 17 55 39.15 | -28 59 53.5 | >12.4-16.3 | Rc | 2451404 | 284 | |
| 512 | 402.47744.10 | 17 55 43.46 | -28 41 01.3 | 10.8:-<14.6 | Rc | 2450965 | 232 | |
| 513 | 402.47744.1236 | 17 55 49.82 | -28 39 09.5 | 14.6-18.4: | Rc | 2451072 | 376 | |
| 514 | 402.47745.178 | 17 55 50.45 | -28 35 42.0 | 15.5:-18.5: | Rc | 2450894 | 333 | |
| 515 | 402.47742.51 | 17 55 51.75 | -28 48 13.4 | >13.2-<17.0 | Rc | 2450918 | 222 | |
| 516 | 402.47796.1591 | 17 56 04.94 | -29 13 21.0 | >14.7-19.3: | Rc | 2450978 | 232 | |
| 517 | 403.47794.99 | 17 56 11.63 | -29 18 35.9 | 12.8-16.9 | Rc | 2451390 | 271 | OGLE-BLG-LPV-127802 |
| 518 | 401.47812.11 | 17 56 12.82 | -28 07 42.3 | 13.9-16.7 | Rc | 2450940 | 402 | OGLE-BLG-LPV-127960 |
| 519 | 401.47870.821 | 17 56 15.15 | -28 16 18.9 | 14.0:-17.2: | Rc | 2450665 | 124 | OGLE-BLG-LPV-128355 |
| 520 | 403.47855.7 | 17 56 16.10 | -29 18 04.4 | >13.2-15.7 | Rc | 2450652 | 209 | |
| 521 | 402.47861.12 | 17 56 16.80 | -28 51 29.4 | 12.8-16.3: | Rc | 2451012 | 206 | |
| 522 | 403.47847.26 | 17 56 17.19 | -29 49 07.8 | 13.0-16.5: | Rc | 2450980 | 258 | OGLE-BLG-LPV-128742 |
| 523 | 403.47844.144 | 17 56 17.84 | -29 59 24.4 | 12.6-16.5 | Rc | 2451253 | 240 | OGLE-BLG-LPV-128864 |
| 524 | 402.47863.43 | 17 56 20.89 | -28 43 40.3 | 12.5-<16.5 | Rc | 2450992 | 240 | |
| 525 | 401.47871.515 | 17 56 21.32 | -28 14 04.0 | 13.6-<18.0 | Rc | 2451370 | 386 | |
| 526 | 403.47853.708 | 17 56 21.91 | -29 25 57.6 | 13.0-16.8 | Rc | 2451343 | 257 | |
| 527 | 402.47865.12 | 17 56 24.28 | -28 38 08.7 | 12.6-<14.8 | Rc | 2451322 | 357 | |
| 528 | 402.47859.92 | 17 56 24.40 | -28 59 25.9 | 12.7-16.4 | Rc | 2451328 | 252 | |
| 529 | 402.47860.1097 | 17 56 26.25 | -28 56 32.4 | 14.5-18.2 | Rc | 2450975 | 173 | OGLE-BLG-LPV-130415 |
| 530 | 403.47849.1260 | 17 56 26.55 | -29 42 14.7 | 12.4-16.7 | Rc | 2450572 | 244 | |
| 531 | 403.47854.30 | 17 56 28.29 | -29 21 40.6 | 12.8-16.3: | Rc | 2450195 | 270 | OGLE-BLG-LPV-130789 |
| 532 | 402.47861.231 | 17 56 30.45 | -28 54 03.5 | 15.4-18.4 | Rc | 2451319 | 216 | OGLE-BLG-LPV-131151 |
| 533 | 402.47864.10 | 17 56 30.91 | -28 39 03.7 | >12.0-<15.6 | Rc | 2451260 | 176 | OGLE-BLG-LPV-131231 |
| 534 | 401.47930.1029 | 17 56 34.85 | -28 17 11.3 | 13.7-17.2 | Rc | 2454248 | 429.5 | OGLE-BLG-LPV-131931, c |
| 535 | 403.47908.117 | 17 56 34.87 | -29 43 55.1 | 11.8-16.4: | Rc | 2451260 | 295 | OGLE-BLG-LPV-131936 |
| 536 | 403.47908.29 | 17 56 37.77 | -29 42 57.1 | 13.5-16.6 | Rc | 2450637 | 428 | OGLE-BLG-LPV-132456 |
| 537 | 402.47920.811 | 17 56 38.88 | -28 56 11.5 | 14.8:-18.0 | Rc | 2450582 | 288 | OGLE-BLG-LPV-132639 |
| 538 | 401.47930.68 | 17 56 39.75 | -28 14 51.3 | 14.0-16.9 | Rc | 2450928 | 248 | OGLE-BLG-LPV-132799 |
| 539 | 402.47915.16 | 17 56 42.12 | -29 16 49.7 | 12.8:-17.0 | Rc | 2450905 | 230 | OGLE-BLG-LPV-133205 |
| 540 | 401.47929.17 | 17 56 42.47 | -28 19 42.9 | 12.4-16.0 | Rc | 2450660 | 201 | OGLE-BLG-LPV-133271 |
| 541 | 403.47904.301 | 17 56 46.56 | -29 59 50.8 | 12.8-16.3 | Rc | 2450972 | 164 | |
| 542 | 403.47904.1888 | 17 56 47.96 | -30 00 00.8 | 15.2-18.5 | Rc | 2451303 | 195 | OGLE-BLG-LPV-134223 |
| 543 | 403.47908.93 | 17 56 49.85 | -29 45 06.2 | 13.7-16.5 | Rc | 2450979 | 272 | |
| 544 | 401.47986.11 | 17 56 50.63 | -28 31 35.3 | 12.5-16.3 | Rc | 2450898 | 236 | OGLE-BLG-LPV-134680 |
| 545 | 403.47973.18 | 17 56 51.01 | -29 25 14.5 | 12.8-16.1: | Rc | 2450560 | 274 | |
| 546 | 401.47988.38 | 17 56 53.20 | -28 23 52.0 | 12.5-16.5 | Rc | 2451361 | 215 | |
| 547 | 403.47971.4455 | 17 56 55.50 | -29 31 19.8 | 16.1-20.2: | V | 2451295 | 288 | V4656 Sgr, OGLE-BLG-LPV-135449 |
| 548 | 403.47964.31 | 17 56 56.12 | -29 59 28.4 | 13.4-16.8 | Rc | 2450705 | 425 | OGLE-BLG-LPV-135555 |
| 549 | 401.47986.30 | 17 56 56.46 | -28 31 25.9 | 14.5-<18.1 | Rc | 2450573 | 171 | OGLE-BLG-LPV-135614 |
| 550 | 401.47996.25 | 17 56 56.77 | -27 53 30.4 | 13.4-18.0 | Rc | 2451313 | 275 | |
| 551 | 401.47991.546 | 17 56 57.06 | -28 14 07.3 | 12.5:-16.1 | Rc | 2451010 | 210 | |
| 552 | 401.47987.82 | 17 56 57.44 | -28 27 11.3 | 12.3-16.4 | Rc | 2450876 | 197 | |
| 553 | 403.47965.28 | 17 56 58.37 | -29 56 37.6 | 12.4-16.0 | Rc | 2450985 | 410 | |
| 554 | 403.47968.172 | 17 57 00.24 | -29 45 45.8 | 13.0-16.2 | Rc | 2451427 | 192 | |
| 555 | 402.47976.33 | 17 57 02.03 | -29 11 18.3 | 12.6-16.7 | Rc | 2451278 | 253 | |
| 556 | 402.47983.526 | 17 57 04.08 | -28 44 15.1 | 14.6-<18.4: | Rc | 2450659 | 366 | OGLE-BLG-LPV-136817 |
| 557 | 402.47976.8 | 17 57 05.26 | -29 10 41.2 | 13.2-16.7 | Rc | 2451414 | 402 | |
| 558 | 401.47991.529 | 17 57 05.93 | -28 13 27.8 | 13.9-17.0 | Rc | 2450938 | 237 | OGLE-BLG-LPV-137086 |
| 559 | 401.47995.786 | 17 57 06.18 | -27 56 17.3 | 14.1-18.0: | Rc | 2450609 | 278 | |
| 560 | 401.47991.606 | 17 57 06.54 | -28 13 14.5 | >13.2-17.0: | Rc | 2451375 | 245 | |
| 561 | 402.48037.33 | 17 57 10.73 | -29 06 41.4 | 13.1-16.4 | Rc | 2451032 | 390 | |
| 562 | 402.48039.31 | 17 57 14.04 | -29 01 05.8 | 13.3-16.4 | Rc | 2451392 | 288 | OGLE-BLG-LPV-138285 |
| 563 | 401.48056.67 | 17 57 16.20 | -27 53 16.6 | 14.6-18.3 | Rc | 2451384 | 239 | |
| 564 | 403.48025.50 | 17 57 16.83 | -29 58 17.3 | 12.5-<16.2 | Rc | 2450958 | 262 | |
| 565 | 402.48041.152 | 17 57 16.83 | -28 53 55.3 | >11.8-15.9 | Rc | 2451430 | 356 | Mis V0531 |
| 566 | 401.48048.56 | 17 57 17.04 | -28 25 57.6 | 14.3-<17.0 | Rc | 2450897 | 232 | OGLE-BLG-LPV-138720 |
| 567 | 401.48049.161 | 17 57 17.36 | -28 22 22.3 | 12.7-16.3 | Rc | 2450895 | 187 | OGLE-BLG-LPV-138761 |
| 568 | 401.48049.1073 | 17 57 17.84 | -28 21 58.4 | 13.5-17.1 | Rc | 2450897 | 282 | OGLE-BLG-LPV-138829 |
| 569 | 401.48054.248 | 17 57 23.93 | -28 01 58.8 | >14.5-17.4 | Rc | 2450870 | 366 | OGLE-BLG-LPV-139727 |
| 570 | 402.48045.455 | 17 57 25.85 | -28 36 04.5 | 14.3:-17.1: | Rc | 2451414 | 325 | OGLE-BLG-LPV-140021 |
| 571 | 401.48114.555 | 17 57 28.35 | -27 58 46.7 | 13.5-17.8 | Rc | 2451014 | 311 | |
| 572 | 401.48112.111 | 17 57 32.62 | -28 10 15.5 | 12.3-16.3 | Rc | 2450676 | 190 | |
| 573 | 402.48096.84 | 17 57 33.70 | -29 11 18.6 | 12.9-16.6 | Rc | 2451410 | 251 | |
| 574 | 401.48115.19 | 17 57 34.67 | -27 55 32.9 | 12.4-16.6: | Rc | 2450157 | 310 | |

```
575  402.48100.1187  17 57 36.49 -28 57 40.4  14.7-17.4:   Rc 2451402  374    OGLE-BLG-LPV-141561
576  402.48096.92    17 57 36.83 -29 10 50.5  14.4-<17.7   Rc 2450626  389    OGLE-BLG-LPV-141618
577  401.48113.23    17 57 37.49 -28 02 46.6  12.2-16.3:   Rc 2450666  377    Mis V0477
578  401.48108.260   17 57 38.27 -28 23 55.0  14.4-18.5:   Rc 2450605  480    OGLE-BLG-LPV-141814
579  401.48107.171   17 57 38.34 -28 29 32.6  13.2:-16.3   Rc 2450531  232
580  401.48106.390   17 57 38.36 -28 31 05.6  13.8-17.1    Rc 2450893  283    OGLE-BLG-LPV-141826
581  401.48116.434   17 57 40.93 -27 51 18.1  16.0-<19.2   Rc 2451403  378
582  401.48174.65    17 57 56.03 -27 58 58.3  13.5-16.7    Rc 2450279  120
583  401.48168.2186  17 57 57.31 -28 25 15.0  12.7-16.0    Rc 2450681  220    Mis V0858
584  401.48167.1504  17 57 57.62 -28 27 30.6  13.9-<17.3   Rc 2450246  272    OGLE-BLG-LPV-144910
585  401.48231.373   17 58 06.33 -28 11 56.4  13.5-17.0    Rc 2451338  404    OGLE-BLG-LPV-146523
586  118.17880.14    17 58 09.41 -30 04 19.2  14.6-18.9    V  2451040  194    V4666 Sgr
587  401.48231.397   17 58 09.97 -28 14 19.8  13.8-17.0    Rc 2451305  256
588  401.48237.16    17 58 11.16 -27 50 00.9  14.0-17.2    Rc 2450650  412
589  401.48235.62    17 58 11.77 -27 56 53.3  15.2-<18.4   Rc 2451288  353.3  OGLE-BLG-LPV-147554, c
590  401.48228.20    17 58 12.16 -28 25 37.2  12.0-15.6    Rc 2450647  242    OGLE-BLG-LPV-147607
591  118.18009.13    17 58 14.39 -30 08 42.5  13.5-<17.5   Rc 2450999  375    OGLE-BLG-LPV-148017
592  401.48235.98    17 58 19.75 -27 57 10.0  15.7-19.3:   Rc 2450634  390    OGLE-BLG-LPV-149029
593  118.18014.144   17 58 20.04 -29 50 11.3  13.4-15.7    Rc 2450932  285    OGLE-BLG-LPV-149084
594  401.48231.348   17 58 20.66 -28 12 00.0  14.0-16.9    Rc 2454231  208.5  OGLE-BLG-LPV-149185, c
595  401.48292.278   17 58 22.92 -28 06 52.3  13.5-<17.2   Rc 2450641  354    OGLE-BLG-LPV-149605
596  401.48294.428   17 58 24.15 -28 01 37.1  13.7-17.5:   Rc 2450572  266    OGLE-BLG-LPV-149841
597  118.18012.472   17 58 24.58 -29 57 58.2  15.1-18.5    Rc 2450582  279    OGLE-BLG-LPV-149937
598  401.48288.405   17 58 24.61 -28 25 06.4  14.5-<18.1   Rc 2451368  363
599  401.48294.763   17 58 24.70 -27 58 57.3  14.5:-17.6   Rc 2450628  265    OGLE-BLG-LPV-149959
600  401.48292.242   17 58 25.89 -28 07 34.0  >14.4-18.1:  Rc 2450542  369
601  401.48287.380   17 58 26.12 -28 26 37.4  14.2-<17.1   Rc 2451358  352
602  401.48293.22    17 58 26.15 -28 05 59.9  >12.3-<16.0  Rc 2450264  233    OGLE-BLG-LPV-150230
603  401.48292.484   17 58 28.03 -28 10 27.4  14.0-18.0:   Rc 2450632  353    OGLE-BLG-LPV-150549
604  401.48296.5     17 58 28.98 -27 52 03.0  13.0-<16.0   Rc 2450649  453    V4668 Sgr
605  401.48292.107   17 58 34.48 -28 08 49.2  14.5-17.8    Rc 2450517  256    OGLE-BLG-LPV-151621
606  401.48291.324   17 58 34.52 -28 13 12.4  13.3-16.7:   Rc 2450573  315    OGLE-BLG-LPV-151636
607  401.48350.14    17 58 39.91 -28 14 52.3  13.2-16.3    Rc 2450614  196    OGLE-BLG-LPV-152533
608  118.18143.140   17 58 47.01 -29 54 38.1  13.8-17.6:   Rc 2450364  415
609  118.18277.190   17 58 49.47 -29 38 14.1  14.1-18.4    Rc 2451392  315
610  118.18270.2512  17 58 54.44 -30 03 52.9  15.8-<20.0:  Rc 2451247  450    OGLE-BLG-LPV-155390
611  401.48410.101   17 59 08.50 -28 17 52.7  12.2-<16.8   Rc 2451034  255
612  113.18421.82    17 59 10.34 -28 42 06.2  12.0-15.1    Rc 2451339  163    OGLE-BLG-LPV-158358
613  401.48409.142   17 59 15.30 -28 21 25.1  14.7-17.6:   Rc 2450633  401    OGLE-BLG-LPV-159315
614  401.48471.587   17 59 22.86 -28 11 41.6  14.2-17.4:   Rc 2450279  256    OGLE-BLG-LPV-160712
615  113.18412.749   17 59 23.46 -29 17 42.2  13.2-<16.9   Rc 2451265  478    OGLE-BLG-LPV-160831
616  401.48471.275   17 59 24.35 -28 12 07.2  13.9-19.0    V  2450274  252    Mis V0641
617  108.18688.5379  17 59 48.97 -28 12 01.2  15.9-22.0:   Rc 2451352  294    OGLE-BLG-LPV-165360
618  108.18684.670   17 59 50.55 -28 31 15.1  14.7-16.9    Rc 2451280  338    OGLE-BLG-LPV-165655
619  118.18659.155   17 59 53.59 -30 08 42.5  15.0-17.7:   Rc 2450520  355    OGLE-BLG-LPV-166141
620  108.18816.57    18 00 04.61 -28 20 45.8  13.6-18.5:   Rc 2451390  125
621  118.18798.395   18 00 12.01 -29 34 52.2  15.2-17.3    Rc 2449988  384
622  176.18827.244   18 00 12.28 -27 37 07.7  15.4-<18.1   Rc 2451055  395    OGLE-BLG-LPV-169132
623  108.18820.1796  18 00 12.48 -28 05 44.5  15.1-19.0:   Rc 2451090  469
624  108.18816.1465  18 00 14.32 -28 19 43.6  15.0-17.2    Rc 2451323  331    OGLE-BLG-LPV-169412
625  108.18819.1271  18 00 18.96 -28 11 25.8  14.7-17.2    Rc 2451383  340
626  108.18816.1283  18 00 20.20 -28 19 55.8  14.1-17.3    Rc 2451270  394    OGLE-BLG-LPV-170300
627  108.18948.438   18 00 25.02 -28 11 48.8  14.8-18.3:   Rc 2450726  413
628  176.18962.250   18 00 33.20 -27 18 12.3  14.1-17.3    Rc 2451335  236
629  108.19082.450   18 00 40.74 -27 57 52.6  13.8-<20.0   Rc 2450708  409    OGLE-BLG-LPV-173320
630  108.19078.3202  18 00 44.74 -28 12 04.9  15.0-19.3    Rc 2451005  418    OGLE-BLG-LPV-173876
631  108.19082.287   18 00 55.31 -27 58 31.1  15.0-<18.2   Rc 2451341  372    OGLE-BLG-LPV-175252
632  108.19211.1360  18 01 10.04 -28 02 27.9  >14.4-18.1   Rc 2450858  372
633  176.19220.1067  18 01 10.06 -27 24 44.8  15.0-18.1:   Rc 2451382  355
634  108.19342.24    18 01 30.19 -27 57 01.4  11.5-15.8    Rc 2449872  237    V4706 Sgr, a
635  176.19352.191   18 01 33.16 -27 18 44.8  15.9-19.5    Rc 2451300  463    OGLE-BLG-LPV-180581
636  108.19468.1211  18 01 39.06 -28 12 19.1  14.1-17.3    Rc 2450715  463    OGLE-BLG-LPV-181343
637  108.19468.207   18 01 39.39 -28 13 54.3  14.2:-17.4   Rc 2451445  343    OGLE-BLG-LPV-181388
638  108.19465.29    18 01 47.39 -28 27 10.9  11.2:-14.0   Rc 2451374  120.5  GSC 06854-00685
639  108.19595.42    18 01 54.64 -28 25 53.8  11.3-15.1    Rc 2453170  195.5  Mis V0546, OGLE-BLG-LPV-183330, a
640  176.19612.498   18 01 54.87 -27 19 32.4  15.7-17.7    Rc 2450658  330    OGLE-BLG-LPV-183355
641  114.19580.126   18 01 56.70 -29 26 37.2  13.1-16.9    Rc 2451433  349
642  113.19583.433   18 01 57.68 -29 12 31.2  14.0-17.5    Rc 2451047  438    OGLE-BLG-LPV-183706
643  114.19582.1686  18 01 58.82 -29 16 36.2  12.7-16.5    Rc 2450871  298    OGLE-BLG-LPV-183862
644  114.19579.17    18 02 02.26 -29 28 53.7  >11.3-15.8   Rc 2450524  296    Mis V0547, OGLE-BLG-LPV-184236
645  108.19592.2243  18 02 02.90 -28 37 03.4  12.5-16.1    Rc 2451276  295    OGLE-BLG-LPV-184313
646  114.19718.14    18 02 20.38 -28 54 13.9  11.9-<14.2   Rc 2450595  235    Mis V0553
647  114.19849.1497  18 02 31.69 -28 48 53.7  12.3-16.0    Rc 2451302  302    OGLE-BLG-LPV-187883
648  114.19844.1893  18 02 33.39 -29 09 37.3  12.0-15.1    Rc 2450608  179
649  114.19970.177   18 02 51.44 -29 24 18.2  >14.0-17.7   Rc 2449048: 357
650  109.19988.32    18 02 57.58 -28 11 51.5  11.2:-15.4   Rc 2451389  223    Mis V0880 OGLE-BLG-LPV-190766
651  104.19991.1317  18 03 01.22 -28 01 46.6  14.0-17.8    Rc 2451453  362    OGLE-BLG-LPV-191136
```

```
652  109.19981.294   18 03 01.78 -28 41 03.9 12.0-16.1   Rc 2451232  283    OGLE-BLG-LPV-191190
653  109.20113.272   18 03 08.17 -28 35 19.4 12.0-16.0   Rc 2451324  234    OGLE-BLG-LPV-191794
654  114.20099.3642  18 03 08.77 -29 29 02.6 15.4-19.8:  Rc 2451015  560
655  114.20105.188   18 03 09.21 -29 05 45.5 12.7-18.5:  Rc 2451408  212    OGLE-BLG-LPV-191899
656  104.20127.996   18 03 10.99 -27 39 23.0 14.4-17.7   Rc 2451248  404    OGLE-BLG-LPV-192094
657  114.20103.1344  18 03 12.30 -29 14 32.4 >16.4-19.8: Rc 2451375  438    OGLE-BLG-LPV-192232
658  109.20113.520   18 03 12.89 -28 35 41.9 12.5-16.4   Rc 2450302  469.0  OGLE-BLG-LPV-192304, c
659  104.20124.1176  18 03 13.84 -27 50 43.5 13.8-17.2   Rc 2451027  203    OGLE-BLG-LPV-192393
660  104.20129.142   18 03 14.53 -27 31 09.3 >13.6-17.0  Rc 2451180: 357    OGLE-BLG-LPV-192453
661  114.20103.1127  18 03 20.66 -29 14 32.3 12.7-17.0   Rc 2450989  424    OGLE-BLG-LPV-193043
662  104.20123.506   18 03 22.28 -27 55 11.4 14.5-18.0:  Rc 2451403  396    OGLE-BLG-LPV-193180
663  104.20259.61    18 03 25.67 -27 31 40.0 12.3-15.8   Rc 2451020  228
664  114.20234.1541  18 03 25.96 -29 11 02.7 12.8-16.0   Rc 2451325  257
665  114.20233.530   18 03 32.51 -29 14 59.1 15.0-<17.4  Rc 2450598  305    OGLE-BLG-LPV-194043
666  104.20252.1432  18 03 41.12 -27 59 33.0 14.6:-17.5  Rc 2451321  290    OGLE-BLG-LPV-194802
667  114.20361.533   18 03 42.01 -29 22 43.7 14.0-<18.4  Rc 2451338  384.6  OGLE-BLG-LPV-194883, c
668  104.20383.1969  18 03 45.48 -27 52 55.4 14.5-18.0   Rc 2451020  433    OGLE-BLG-LPV-195147
669  104.20384.290   18 03 47.15 -27 48 08.2 >13.0-16.5  Rc 2451440  427
670  104.20384.732   18 03 50.09 -27 49 19.6 14.4-17.5   Rc 2450552  303
671  104.20380.3404  18 03 56.72 -28 04 29.5 >13.3-15.9  Rc 2451257  408    OGLE-BLG-LPV-196119
672  104.20388.795   18 03 56.97 -27 31 54.2 13.5-16.9   Rc 2451417  343
673  104.20383.48    18 03 58.22 -27 53 37.6 11.6-16.0   Rc 2451250  218    OGLE-BLG-LPV-196281
674  104.20519.21    18 04 03.22 -27 30 43.4 11.4:-15.3  Rc 2451305  240
675  114.20492.62    18 04 08.95 -29 19 26.8 >12.0-14.9  Rc 2450998  246    OGLE-BLG-LPV-197307
676  109.20508.87    18 04 11.53 -28 14 59.1 11.7-15.7   Rc 2451312  251
677  114.20499.1439  18 04 15.30 -28 51 00.4 10.8-14.7   Rc 2451343  335
678  104.20640.4674  18 04 19.73 -28 04 55.0 11.2:-14.5  Rc 2451296  190
679  104.20649.5962  18 04 20.41 -27 28 17.6 11.7:-14.7  Rc 2451380  114.5
680  114.20620.819   18 04 31.45 -29 24 40.2 14.0-17.2   Rc 2451062  420    OGLE-BLG-LPV-199537
681  101.20654.170   18 04 33.00 -27 08 20.8 13.1-16.6   Rc 2451230  365    OGLE-BLG-LPV-199690
682  101.20787.65    18 04 37.83 -26 58 57.7 13.3-16.4   Rc 2449997  401    OGLE-BLG-LPV-200174
683  101.20785.33    18 04 38.33 -27 07 26.6 11.8:-15.4  Rc 2451385  324    OGLE-BLG-LPV-200228
684  101.20786.1066  18 04 40.88 -27 00 36.7 12.4-17.1   Rc 2451468  360
685  101.20779.191   18 04 51.24 -27 28 37.3 >12.8-15.7: Rc 2450175  322
686  114.20883.41    18 04 56.67 -29 14 22.0 11.0:-15.2  Rc 2451342  228    OGLE-BLG-LPV-202245
687  109.20891.113   18 04 57.89 -28 43 28.1 11.9-15.3   Rc 2451363  353
688  104.21032.445   18 05 28.73 -27 57 55.3 >13.0-16.1  Rc 2451364  296    OGLE-BLG-LPV-205560
689  101.21173.381   18 05 32.95 -27 14 03.9 14.5-<17.7  Rc 2451437  397
690  101.21176.964   18 05 34.24 -27 03 28.8 >13.5-17.7  Rc 2451388  339
691  101.21170.797   18 05 43.68 -27 25 15.9 13.9-16.9:  Rc 2451262  380    OGLE-BLG-LPV-206825
692  120.21269.13    18 05 58.23 -29 31 27.9 10.8-14.8   Rc 2451337  310    a
693  101.21307.439   18 06 00.41 -26 56 24.5 14.0:-17.2  Rc 2450865  375
694  101.21429.60    18 06 10.17 -27 30 20.9 11.1:-15.5  Rc 2450167  318
695  128.21407.362   18 06 10.42 -28 58 42.1 12.7-15.8   Rc 2451313  399
696  101.21437.204   18 06 12.44 -26 59 16.9 11.6-<15.8  Rc 2450628  443
697  128.21409.540   18 06 15.09 -28 49 19.4 13.2-17.0   Rc 2451370  345
698  128.21409.579   18 06 16.38 -28 49 50.7 15.3-20.4:  Rc 2451028  483
699  101.21428.114   18 06 17.90 -27 33 31.1 >12.7-15.8  Rc 2451213  358
700  101.21432.307   18 06 21.40 -27 16 22.4 12.5-16.6   Rc 2451440  346    OGLE-BLG-LPV-209515
701  120.21395.990   18 06 22.53 -29 46 23.3 13.2-16.9:  Rc 2451289  324    OGLE-BLG-LPV-209583
702  120.21400.65    18 06 24.21 -29 24 42.1 10.8-14.6   Rc 2451390  195.5 a
703  128.21538.224   18 06 28.43 -28 52 56.3 >14.5-16.9  Rc 2450328  433
704  128.21541.15    18 06 28.45 -28 40 58.7 12.5-16.4   Rc 2450536  289    OGLE-BLG-LPV-209889
705  128.21534.130   18 06 29.26 -29 07 53.9 14.1-16.9   Rc 2450162  349    OGLE-BLG-LPV-209932
706  179.21579.922   18 06 33.47 -26 10 09.7 15.3-18.9   Rc 2451403  340    OGLE-BLG-LPV-210173
707  179.21585.69    18 06 33.48 -25 46 38.1 13.4:-17.0: Rc 2453557  442.8  OGLE-BLG-LPV-210174, c
708  179.21585.2061  18 06 33.61 -25 46 50.3 15.1-20.0:  Rc 2451375  342    OGLE-BLG-LPV-210185
709  105.21552.3440  18 06 43.06 -27 58 54.7 11.9-<16.6  Rc 2450534  307
710  120.21655.785   18 06 44.42 -29 45 33.0 15.7-18.7   Rc 2450200  518
711  105.21684.924   18 06 45.92 -27 49 44.0 14.2-17.7   Rc 2451222  405    OGLE-BLG-LPV-210891
712  179.21705.30    18 06 48.92 -26 27 34.0 11.0-<14.1  Rc 2451007  450
713  128.21672.319   18 06 54.62 -28 37 24.5 12.4-15.8   Rc 2451392  305    OGLE-BLG-LPV-211391
714  105.21678.336   18 06 55.43 -28 13 58.8 11.7-15.9   Rc 2450711  356
715  105.21683.366   18 06 58.83 -27 54 47.3 >15.6-17.8: Rc 2450973  456    OGLE-BLG-LPV-211621
716  128.21793.262   18 07 10.33 -29 12 15.4 >14.8-17.8  Rc 2450970  502
717  105.21814.12    18 07 17.16 -27 50 41.4 >12.7-16.4  Rc 2449790  339
718  120.21787.672   18 07 19.56 -29 37 45.9 14.4-17.0   Rc 2451390  435
719  128.21934.68    18 07 23.83 -28 31 38.6 11.6-<15.9  Rc 2450270  252
720  179.21972.669   18 07 33.59 -25 57 52.1 14.2-17.9   Rc 2451332  308    OGLE-BLG-LPV-213860
721  128.22055.126   18 07 40.13 -29 06 36.3 >13.5-17.4  Rc 2450110  353    OGLE-BLG-LPV-214321, c
722  105.22073.52    18 07 40.18 -27 53 53.4 13.8-<18.5  Rc 2450615  500    OGLE-BLG-LPV-214324
723  128.22056.213   18 07 41.31 -29 03 21.4 14.0-16.3   Rc 2450726  371
724  128.22061.568   18 07 41.33 -28 43 21.8 14.6-18.7:  Rc 2450715  195    OGLE-BLG-LPV-214394
725  105.22071.39    18 07 43.96 -28 03 31.5 >11.9-15.6  Rc 2449247  384
726  128.22058.402   18 07 46.49 -28 52 44.0 14.5-17.2   Rc 2451245  444    OGLE-BLG-LPV-214720
727  180.22111.49    18 07 49.76 -25 20 50.9 14.4-<17.5  Rc 2452426  365.0  OGLE-BLG-LPV-214920, c
728  180.22109.195   18 07 50.55 -25 28 26.9 >14.0-16.8  Rc 2450924  305
```

```
729  128.22187.3833  18 07 57.28 -28 59 34.1 10.8:-<15.0 Rc 2450561  191.5 ASAS J180759-2859.5
730  180.22247.288   18 08 01.29 -24 59 29.0 15.3-<19.1  Rc 2450678  376
731  180.22246.22    18 08 04.69 -24 59 52.6 12.3-16.3   Rc 2450925  244
732  110.22193.126   18 08 09.89 -28 31 42.4 12.8-16.0   Rc 2451343  290
733  110.22194.156   18 08 10.33 -28 31 32.2 >14.5-17.2  Rc 2450125  362   OGLE-BLG-LPV-216140
734  128.22313.1871  18 08 16.09 -29 11 48.9 12.9-17.6   Rc 2451372  337   OGLE-BLG-LPV-216471
735  180.22376.685   18 08 21.34 -24 59 45.6 13.9-<18.0  Rc 2451363  346
736  110.22323.31    18 08 25.59 -28 32 20.6 11.8:-<17.0 Rc 2450214  260.3 a
737  180.22371.729   18 08 28.71 -25 22 36.9 14.8-<18.0  Rc 2451335  400
738  110.22320.1968  18 08 29.17 -28 46 55.1 11.4:-<15.1 Rc 2451056  292   a
739  180.22371.54    18 08 29.49 -25 20 09.8 13.3-15.9   Rc 2451249  316
740  179.22356.1295  18 08 29.80 -26 22 09.6 15.2:-<19.5 Rc 2450627  203   OGLE-BLG-LPV-217242
741  115.22437.1157  18 08 34.34 -29 38 18.5 15.2-17.8   Rc 2451292  406   OGLE-BLG-LPV-217485
742  110.22453.546   18 08 38.16 -28 32 22.9 14.5-18.1   Rc 2451344  181   OGLE-BLG-LPV-217707
743  110.22454.501   18 08 41.26 -28 28 05.8 >13.3-16.9  Rc 2450785: 364
744  102.22463.4     18 08 41.83 -27 54 06.7 12.0-15.9   Rc 2451418  285
745  110.22449.2753  18 08 43.76 -28 48 45.7 14.8-19.0   Rc 2451024  546
746  102.22598.36    18 08 54.67 -27 32 08.2 >11.7-15.5  Rc 2449054  347
747  180.22630.134   18 08 54.74 -25 27 16.0 13.4-15.4   Rc 2451285  400
748  110.22583.125   18 08 55.36 -28 33 07.0 14.5:-17.6  Rc 2451262  234
749  110.22579.37    18 08 55.81 -28 50 20.0 11.9-14.7   Rc 2451422  180
750  110.22585.827   18 08 57.41 -28 25 12.7 14.0-17.9   Rc 2450284  260   OGLE-BLG-LPV-218704
751  180.22633.82    18 08 58.33 -25 14 12.6 14.6-<16.5  Rc 2451340  374
752  102.22591.90    18 08 58.85 -28 02 17.7 11.8-15.3:  Rc 2451403  367
753  110.22576.29    18 08 59.15 -29 03 15.3 11.0:-14.8  Rc 2451263  231
754  115.22563.56    18 08 59.66 -29 52 08.9 13.2-<15.7  Rc 2451312  280
755  180.22639.364   18 09 06.34 -24 50 31.8 14.1-18.5:  Rc 2451355  418
756  102.22724.2834  18 09 13.48 -27 48 39.7 15.4-18.8   Rc 2451288  187.5 OGLE-BLG-LPV-219517
757  115.22699.120   18 09 13.77 -29 28 04.4 11.6-15.3   Rc 2450633  270
758  180.22767.528   18 09 16.18 -24 58 34.6 14.8-18.0   Rc 2451270  453
759  178.22747.118   18 09 16.32 -26 18 53.2 12.7-17.4   Rc 2451335  305
760  110.22708.113   18 09 18.83 -28 54 45.2 11.7-<15.0  Rc 2451275  254
761  110.22705.6     18 09 20.74 -29 05 08.5 >11.8-<15.3 Rc 2451395  467
762  102.22722.585   18 09 21.05 -27 59 03.2 14.4-17.2   Rc 2451362  336   OGLE-BLG-LPV-219838
763  110.22712.317   18 09 21.93 -28 38 02.4 14.4-17.7   Rc 2450330  445
764  180.22769.425   18 09 22.44 -24 48 19.7 15.3-18.5:  Rc 2450685  404
765  180.22759.4919  18 09 25.48 -25 30 44.8 14.3-17.8:  Rc 2451420  161
766  115.22701.62    18 09 27.23 -29 22 58.7 12.9-<15.5  Rc 2451268  346   OGLE-BLG-LPV-220116
767  115.22700.113   18 09 27.28 -29 25 15.2 >13.4-<16.8 Rc 2451247: 363   OGLE-BLG-LPV-220118
768  110.22708.91    18 09 27.43 -28 54 50.7 12.6-16.0   Rc 2451335  135
769  115.22696.162   18 09 27.63 -29 40 13.8 12.0-<15.8  Rc 2450198  304
770  115.22699.890   18 09 27.98 -29 30 05.1 >13.8-17.9  Rc 2451290: 395   OGLE-BLG-LPV-220166
771  180.22890.135   18 09 35.34 -25 24 24.3 14.0-<16.7  Rc 2452489  381.1 OGLE-BLG-LPV-220499, c
772  115.22832.17    18 09 37.69 -29 19 21.5 12.9-18.0   Rc 2451307  310   OGLE-BLG-LPV-220609
773  102.22851.229   18 09 38.27 -28 03 27.4 12.2:-16.2  Rc 2451265  378
774  102.22854.543   18 09 38.69 -27 47 41.5 13.7-16.4   Rc 2450649  428
775  102.22859.31    18 09 39.15 -27 30 09.6 11.0:-15.4  Rc 2454524  267   a
776  180.22896.764   18 09 42.78 -25 02 47.9 15.0-18.6   Rc 2451289  190.5 OGLE-BLG-LPV-220814
777  110.22975.111   18 09 46.63 -28 25 51.5 14.7-<16.3  Rc 2450598  373   OGLE-BLG-LPV-220954
778  115.22955.133   18 09 46.88 -29 45 23.2 13.0-16.2   Rc 2450304  443   OGLE-BLG-LPV-220966
779  110.22975.402   18 09 49.42 -28 26 51.6 12.2-16.4   Rc 2450695  359
780  115.22956.161   18 09 50.82 -29 40 27.0 12.8-16.3   Rc 2450560  310
781  102.22988.46    18 09 53.69 -27 35 31.7 11.5-15.0   Rc 2450551  274
782  110.22966.1853  18 09 55.18 -29 01 25.7 15.0-<21.0: Rc 2450534  554
783  110.22969.81    18 09 56.15 -28 49 39.6 >11.7-14.9  Rc 2451283  472
784  180.23025.69    18 09 56.96 -25 06 58.1 10.6-17.4   Rc 2449902  443.4 OGLE-BLG-LPV-221392, c
785  110.22970.1618  18 09 57.36 -28 47 04.8 >13.7-17.2  Rc 2450162  329
786  102.22985.728   18 10 00.56 -27 45 05.5 12.8-17.0   Rc 2451373  343
787  180.23029.75    18 10 01.54 -24 50 32.6 15.3-18.8   Rc 2449844  327
788  180.23154.43    18 10 05.54 -25 08 24.8 15.3-<17.8  Rc 2450607  425   OGLE-BLG-LPV-221744
789  102.23120.358   18 10 10.81 -27 26 41.6 12.4-16.7   Rc 2450643  490
790  115.23083.24    18 10 13.13 -29 52 42.8 12.7-17.0   Rc 2451315  320   OGLE-BLG-LPV-222028
791  102.23110.83    18 10 13.94 -28 04 21.3 13.4-15.7   Rc 2451356  296
792  102.23117.494   18 10 14.35 -27 36 25.2 13.1:-16.3  Rc 2449067  358
793  115.23214.146   18 10 27.79 -29 51 03.4 12.0-15.6   Rc 2451252  248
794  102.23243.363   18 10 29.08 -27 53 16.5 14.4-17.0   Rc 2450277  414   OGLE-BLG-LPV-222645
795  115.23213.1551  18 10 31.71 -29 52 24.4 15.0-19.0:  Rc 2450545  447   OGLE-BLG-LPV-222753
796  115.23213.20    18 10 32.36 -29 53 27.0 10.2:-<14.2 Rc 2452901  192   a
797  115.23213.17    18 10 35.84 -29 52 32.8 >11.7-<15.1 Rc 2449125  350
798  102.23250.264   18 10 36.95 -27 24 24.0 14.8-17.5:  Rc 2450657  295   OGLE-BLG-LPV-222951
799  110.23227.77    18 10 38.46 -28 59 07.9 13.0-<15.8  Rc 2451362  378
800  102.23240.22    18 10 38.74 -28 05 04.1 12.2-<15.8  Rc 2450680  489
801  110.23226.21    18 10 39.80 -28 59 51.1 11.5-14.7   Rc 2450975  200
802  115.23215.70    18 10 40.24 -29 47 30.0 >11.8-14.7  Rc 2451375  214
803  110.23359.47    18 10 42.27 -28 48 41.0 >13.5-16.4  Rc 2449127  331
804  180.23412.6     18 10 42.35 -25 19 36.2 11.7-14.2   Rc 2450569  152
805  102.23372.281   18 10 47.06 -27 56 15.2 11.5:-16.0  Rc 2450652  275   a
```

```
806  102.23376.436  18 10 49.27 -27 41 19.7 11.7:-16.7   Rc 2451443  269
807  102.23373.76   18 10 54.17 -27 54 21.3 13.0-16.2    Rc 2450657  495
808  102.23511.23   18 11 03.84 -27 22 34.7 13.5-<16.1   Rc 2450305  393
809  102.23510.1015 18 11 04.85 -27 24 27.9 14.1-18.0    Rc 2450572  431.1 OGLE-BLG-LPV-223907, c
810  115.23473.377  18 11 06.95 -29 52 55.5 14.3-<17.4   Rc 2451020  352
811  111.23492.5069 18 11 17.05 -28 37 23.1 11.7:-14.0   Rc 2450614  262   a
812  111.23483.459  18 11 17.32 -29 15 40.7 13.9-17.5    Rc 2450285  458
813  102.23634.99   18 11 22.57 -27 49 47.6 13.7-<15.5   Rc 2450572  347
814  102.23634.254  18 11 26.56 -27 50 10.0 15.0-17.5:   Rc 2451407  228
815  178.23788.141  18 11 42.73 -26 14 01.2 13.0-15.7    Rc 2451043  474
816  102.23764.71   18 11 49.68 -27 49 02.2 >11.7-14.1   Rc 2451247  385
817  167.23784.190  18 11 54.28 -26 30 00.1 13.9-16.6    Rc 2453248  233.4 OGLE-BLG-LPV-225157, c
818  103.24034.3927 18 12 14.31 -27 11 39.1 12.8-16.2    Rc 2451320  486
819  103.24028.193  18 12 16.53 -27 35 29.6 13.3-<15.7   Rc 2450564  310
820  103.24159.188  18 12 32.43 -27 28 52.1 13.5-17.5:   Rc 2449088  419
821  111.24135.446  18 12 33.49 -29 04 13.9 >13.5-17.4   Rc 2451326  194
822  103.24155.31   18 12 38.18 -27 44 58.2 11.1-15.0    Rc 2451379  326
823  103.24159.422  18 12 43.44 -27 28 30.6 13.2-16.9    Rc 2451372  422
824  111.24136.3779 18 12 46.75 -29 02 39.1 11.5-14.6    Rc 2450190  313
825  111.24137.4010 18 12 48.13 -28 58 50.2 11.3-15.7    Rc 2451365  305
826  111.24268.35   18 12 50.65 -28 51 45.5 >11.0-15.2   Rc 2451415: 345
827  103.24288.1033 18 12 55.53 -27 32 38.9 14.9-17.7    Rc 2451010  439
828  111.24268.83   18 12 55.91 -28 53 51.9 >11.5-15.0   Rc 2451295: 400
829  103.24287.464  18 13 00.57 -27 36 25.0 12.9-17.3    Rc 2451055  550   NSV 10359
830  111.24268.46   18 13 02.01 -28 52 04.0 >11.4-14.4   Rc 2451337  303
831  111.24264.332  18 13 03.21 -29 08 20.5 >12.8-<16.4  Rc 2451060: 363
832  111.24270.80   18 13 03.61 -28 47 31.4 11.2-14.4    Rc 2450608  229
833  103.24291.654  18 13 03.69 -27 21 23.6 14.6-<17.1   Rc 2450583  315
834  103.24413.137  18 13 14.38 -27 51 46.6 12.3-17.1    Rc 2450672  421
835  103.24417.536  18 13 16.78 -27 36 37.4 >13.2:-16.9  Rc 2451210: 371
836  116.24387.51   18 13 19.79 -29 36 19.1 12.7-<17.2   Rc 2450273  376   OGLE-BLG-LPV-226558
837  103.24419.126  18 13 24.88 -27 28 49.3 >12.9-<16.6  Rc 2450287  255
838  103.24415.40   18 13 24.99 -27 46 07.6 >13.0-16.9:  Rc 2449192  380
839  111.24524.305  18 13 32.22 -29 07 54.9 >12.1-15.8   Rc 2450621  276
840  103.24547.1405 18 13 37.36 -27 36 04.3 11.5-14.8    Rc 2451301  284
841  103.24552.61   18 13 37.74 -27 17 49.9 12.4-15.7:   Rc 2449083  350
842  111.24522.18   18 13 40.29 -29 15 54.5 10.6-14.8    Rc 2449209  266   a
843  103.24683.102  18 13 50.99 -27 14 00.3 13.2-15.8:   Rc 2451343  300   OGLE-BLG-LPV-227045
844  111.24787.377  18 14 02.51 -28 57 58.7 >13.5-16.3:  Rc 2449856: 311
845  161.24829.65   18 14 03.60 -26 08 14.6 13.8-15.5:   Rc 2450907  437
846  103.24805.4730 18 14 14.18 -27 44 07.3 15.3-18.2:   Rc 2451000  439
847  103.24809.3394 18 14 18.01 -27 31 01.7 14.0-17.3    Rc 2450633  216
848  161.24956.149  18 14 28.20 -26 22 00.3 14.7-16.9    Rc 2450218  261
849  307.35038.26   18 14 28.60 -23 52 05.0 12.2-16.5    Rc 2451318  276
850  306.35051.162  18 14 33.35 -22 59 08.4 14.0-17.1    Rc 2450936  236
851  304.35058.101  18 14 34.21 -22 33 04.2 >13.9-17.3   Rc 2451230: 370
852  306.35055.57   18 14 34.91 -22 46 48.0 13.3-17.0    Rc 2451015  447
853  305.35067.40   18 14 35.04 -21 56 00.3 15.4-19.3    Rc 2450537  309
854  167.24948.31   18 14 35.65 -26 55 37.3 10.7-14.1:   Rc 2450905  171   a
855  306.35214.385  18 14 37.00 -23 22 00.2 14.2-18.0    Rc 2451352  346
856  306.35214.120  18 14 37.21 -23 22 45.6 13.3-16.9    Rc 2450578  270
857  305.35238.56   18 14 40.97 -21 45 43.8 14.5-<17.8   Rc 2451287  352
858  305.35236.114  18 14 44.16 -21 53 41.4 15.4-19.0    Rc 2450202  316
859  307.35203.9    18 14 45.83 -24 05 19.1 13.1-15.8    Rc 2450313  389
860  307.35208.632  18 14 49.62 -23 43 21.3 13.9-21.5:   Rc 2450595  283
861  306.35390.48   18 14 53.43 -22 47 03.6 13.5-<16.9   Rc 2451319  296
862  304.35391.339  18 14 54.73 -22 43 11.1 15.0-18.0    Rc 2451297  400
863  305.35411.149  18 14 55.18 -21 23 43.8 14.4:-<17.2  Rc 2450955  447
864  304.35400.159  18 14 58.63 -22 10 09.2 15.4-<18.4   Rc 2451352  372
865  305.35412.379  18 15 01.99 -21 22 20.1 16.5-20.2:   Rc 2450530  194
866  307.35373.2725 18 15 04.19 -23 56 47.9 >14.0-20.0:  Rc 2451303  400
867  305.35410.181  18 15 04.66 -21 28 23.1 >15.5-19.4:  Rc 2450645  425
868  307.35376.82   18 15 06.16 -23 44 30.1 14.3:-<17.3  Rc 2451402  370
869  305.35402.184  18 15 06.26 -22 01 33.1 >14.5-17.5   Rc 2450470  358
870  304.35399.174  18 15 06.76 -22 14 44.8 >14.3-17.0   Rc 2451185: 359
871  304.35567.31   18 15 11.15 -22 13 29.5 13.1-<16.0   Rc 2451368  385
872  307.35541.147  18 15 14.33 -23 58 10.4 12.9-16.0    Rc 2450961  300
873  306.35556.110  18 15 15.60 -22 58 44.4 12.5-16.4    Rc 2451305  268.5
874  307.35548.86   18 15 17.62 -23 29 00.7 14.0-17.2    Rc 2450194  270
875  177.25361.49   18 15 21.49 -25 20 08.9 14.4-18.7:   Rc 2450960  259   OGLE-BLG-LPV-228268
876  305.35576.146  18 15 22.06 -21 36 37.9 14.7-17.6    Rc 2449990: 400
877  306.35555.283  18 15 23.54 -23 01 22.8 14.7-<18.5   Rc 2450594  409
878  306.35718.463  18 15 26.54 -23 21 48.3 >14.2-<17.1  Rc 2450230  238
879  305.35738.316  18 15 28.86 -22 00 23.9 14.0-17.4    Rc 2450956  290
880  304.35729.66   18 15 33.48 -22 37 28.3 12.8-16.5    Rc 2451347  405
881  305.35745.791  18 15 34.01 -21 32 32.8 14.8-18.4    Rc 2451390  343
882  306.35723.119  18 15 34.44 -23 02 42.4 14.2-16.8    Rc 2450239  272
```

```
883  304.35728.128   18 15 35.50  -22 40 20.6  14.9-<18.0   Rc 2451368   372
884  306.35726.165   18 15 35.9   -22 49 59.4  14.6-18.4    Rc 2450184   344
885  305.35745.91    18 15 36.22  -21 34 39.1  >13.6-16.3   Rc 2449750:  343
886  177.25498.648   18 15 36.47  -24 54 53.8  16.1-19.1    Rc 2451326   459   OGLE-BLG-LPV-228414
887  305.35744.432   18 15 38.30  -21 37 48.3  14.1-17.3    Rc 2451383   335
888  306.35718.73    18 15 39.83  -23 21 34.2  12.6-15.7    Rc 2450306   317
889  304.35736.146   18 15 41.13  -22 07 57.0  14.3-16.8    Rc 2449807   327
890  305.35911.758   18 15 43.36  -21 41 34.4  16.7-20.5:   Rc 2450513   552
891  157.25386.28    18 15 45.70  -32 20 02.6  >11.6-14.6   Rc 2449630:  373
892  305.35907.164   18 15 49.11  -21 57 52.8  13.0-17.6    Rc 2451021   488
893  305.35913.104   18 15 50.34  -21 33 49.5  13.5-16.2    Rc 2450670   147
894  306.35895.352   18 15 52.36  -22 46 33.1  14.8-18.8    Rc 2451387   350
895  304.35896.13    18 15 53.39  -22 41 48.1  13.0-<17.6   Rc 2451350   369
896  306.35888.85    18 15 56.02  -23 14 40.7  13.0-16.7    Rc 2450539   302
897  305.35915.484   18 15 56.80  -21 26 44.7  14.5-17.1    Rc 2451340   437
898  305.35914.127   18 15 57.05  -21 29 43.6  16.1-19.7    Rc 2450307   418
899  304.36063.242   18 16 02.83  -22 43 59.8  >14.1-17.5   Rc 2450002:  376
900  305.36080.175   18 16 03.01  -21 38 25.4  13.8:-17.0   Rc 2450880   384
901  157.25513.177   18 16 07.36  -32 33 15.1  11.8:-<16.4  Rc 2451355   356
902  306.36060.195   18 16 08.66  -22 55 19.9  >14.3-17.5   Rc 2450530   367
903  304.36063.337   18 16 12.35  -22 44 06.1  15.3-18.1    Rc 2450545   241
904  305.36076.320   18 16 16.57  -21 53 21.8  14.4-19.0    Rc 2450577   412
905  306.36230.10    18 16 18.20  -22 49 21.3  >12.4-16.3   Rc 2450194   302
906  305.36242.66    18 16 19.28  -22 00 12.7  12.9-16.5    Rc 2451315   273
907  306.36226.238   18 16 23.33  -23 03 15.8  13.8-16.8    Rc 2450642   292
908  305.36250.400   18 16 26.32  -21 28 11.0  14.0-17.4:   Rc 2450581   203
909  306.36221.243   18 16 26.44  -23 25 11.1  12.6-16.9    Rc 2450945   234
910  304.36232.26    18 16 31.20  -22 42 54.9  >11.5-14.3   Rc 2451100   430
911  305.36414.88    18 16 42.96  -21 46 29.6  13.3-<17.1   Rc 2450886   363
912  306.36557.77    18 16 55.99  -23 23 12.0  12.4-15.4    Rc 2451268   317
913  305.36579.607   18 16 57.59  -21 58 54.8  14.3-<18.3   Rc 2450584   316
914  305.36579.19    18 16 57.61  -21 56 09.7  12.8-<16.6   Rc 2451267   337
915  306.36561.35    18 16 59.54  -23 10 34.1  >13.0-15.3:  Rc 2450983   221
916  157.25904.24    18 17 04.51  -32 29 05.3  >12.7-15.1   Rc 2449750:  362
917  155.26045.67    18 17 09.66  -31 44 21.4  >13.0-16.4   Rc 2451376   316
918  305.36917.497   18 17 26.96  -21 47 31.3  14.8-18.0    Rc 2451245:  402
919  304.36906.128   18 17 27.67  -22 31 42.7  14.3-<16.7   Rc 2451352   327
920  305.36919.1000  18 17 27.85  -21 42 45.0  15.7-<19.2   Rc 2450664   467
921  305.36917.36    18 17 30.51  -21 47 41.4  12.7-15.6    Rc 2450978   168.5
922  305.36918.44    18 17 31.51  -21 46 16.0  11.8-15.5    Rc 2450542   280
923  309.36911.41    18 17 32.72  -22 11 12.7  11.5-15.5    Rc 2450610   215
924  308.36917.519   18 17 35.07  -21 47 12.7  13.3-17.5    Rc 2450592   283
925  308.36915.105   18 17 35.40  -21 58 13.4  12.5-16.4    Rc 2450891   233
926  308.37090.146   18 17 42.45  -21 28 09.3  13.1-16.7    Rc 2450925   290
927  308.37081.19    18 17 42.46  -22 03 11.5  12.4-16.6    Rc 2450233   273
928  308.37090.433   18 17 46.63  -21 27 15.1  14.0-18.0    Rc 2450932   404
929  309.37076.962   18 17 48.93  -22 26 07.2  14.5-17.6    Rc 2450540   356
930  308.37087.4     18 17 55.44  -21 39 51.2  12.2-16.0    Rc 2450924   255
931  308.37084.235   18 17 56.03  -21 54 26.1  13.8:-17.2   Rc 2450894   365
932  311.37227.694   18 17 58.70  -23 31 49.5  14.5-18.3    Rc 2451297   383
933  309.37247.217   18 18 01.63  -22 13 10.2  12.7-16.7    Rc 2451346   337
934  310.37236.71    18 18 02.97  -22 55 56.7  12.1-16.6    Rc 2451330   234
935  309.37244.9     18 18 05.00  -22 23 44.3  13.6-17.2    Rc 2449876   278
936  155.26430.26    18 18 05.87  -32 03 54.6  >12.1-15.6   Rc 2449630:  383
937  310.37231.57    18 18 05.96  -23 15 11.8  12.3-16.4    Rc 2451325   232
938  308.37259.119   18 18 07.30  -21 23 12.8  13.5-17.0    Rc 2450540   298
939  311.37223.1206  18 18 08.88  -23 49 11.0  13.4-17.0:   Rc 2450535   331
940  310.37237.368   18 18 08.93  -22 52 40.6  13.2-17.5    Rc 2451294   398
941  308.37252.26    18 18 10.25  -21 54 08.7  12.3-16.5    Rc 2451340   233
942  311.37225.124   18 18 14.43  -23 42 31.6  >13.2-16.8:  Rc 2451281   382
943  311.37227.89    18 18 14.45  -23 34 12.0  13.8:-16.3:  Rc 2450637   428
944  309.37247.71    18 18 14.88  -22 12 46.2  12.6-16.6    Rc 2450968   295
945  310.37407.177   18 18 16.38  -22 46 48.9  13.5-17.2    Rc 2450539   283
946  155.26431.24    18 18 17.80  -32 01 55.6  13.2-15.6    Rc 2451301   315
947  309.37407.243   18 18 19.35  -22 44 35.2  14.5-16.9    Rc 2451308   309
948  308.37421.123   18 18 21.12  -21 48 56.0  14.6-18.3    Rc 2451334   511
949  310.37399.459   18 18 22.59  -23 18 27.7  13.6-17.6    Rc 2450977   475
950  308.37417.47    18 18 23.60  -22 03 13.7  15.2-19.2    Rc 2450941   222
951  311.37394.18    18 18 28.32  -23 35 32.4  12.2-16.6    Rc 2451279   138
952  308.37425.44    18 18 30.61  -21 34 24.5  13-<16.6     Rc 2450221   290
953  310.37401.23    18 18 30.71  -23 10 34.4  12.5-16.2:   Rc 2451403   253
954  155.26565.69    18 18 34.28  -31 46 03.8  12.5-16.4    Rc 2450951   156
955  309.37576.222   18 18 39.42  -22 41 50.5  13.9-<17.1   Rc 2450580   298
956  310.37572.20    18 18 40.57  -22 57 47.5  >13.8-16.9   Rc 2450205   308
957  311.37564.351   18 18 43.55  -23 29 51.2  13.8-17.5    Rc 2451294   328
958  308.37585.95    18 18 43.58  -22 03 48.9  13.6:-<17.8  Rc 2450587   287
959  310.37573.9     18 18 45.13  -22 54 56.6  12.4-15.7    Rc 2450196   311
```

```
960  310.37740.758   18 18 56.11 -22 57 34.0 12.1-<16.0  Rc 2450670  220
961  308.37762.2492  18 19 04.10 -21 30 55.7 >12.4-<16.3 Rc 2451330  300
962  160.27048.224   18 19 23.87 -25 34 43.4 14.6:-19.4  Rc 2450680: 559
963  310.38079.25    18 19 27.25 -22 46 57.0 12.4-<16.4  Rc 2450168  362
964  310.38077.52    18 19 33.17 -22 53 40.7 14.1-17.4   Rc 2450266  195
965  309.38088.127   18 19 34.13 -22 07 08.7 14.2-18.0:  Rc 2451333  299
966  149.27108.65    18 19 39.32 -30 13 28.3 >11.7-15.0  Rc 2449899  304
967  308.38266.162   18 19 42.61 -21 30 31.3 13.0-17.5   Rc 2450642  314
968  310.38242.89    18 19 44.17 -23 04 54.6 13.8-16.6   Rc 2450560  406
969  311.38230.1146  18 19 46.27 -23 52 05.7 13.6-17.9   Rc 2450594  534
970  311.38398.126   18 19 59.30 -23 54 14.3 11.5:-15.5  Rc 2451272  257
971  149.27240.6     18 20 01.31 -30 07 24.5 11.6-14.6   Rc 2451295  175.5
972  311.38567.1070  18 20 14.38 -23 48 29.2 12.5-15.8   Rc 2451283  380
973  308.38597.243   18 20 21.20 -21 47 14.8 13.3-17.1   Rc 2450574  263
974  149.27360.929   18 20 22.02 -30 46 55.1 >13.0-15.8  Rc 2451300  402
975  310.38577.136   18 20 24.52 -23 08 49.8 12.3-16.9:  Rc 2450679  523
976  308.38597.974   18 20 24.60 -21 50 53.6 15.2-18.8   Rc 2450970  320
977  149.27752.4254  18 21 16.74 -30 36 47.5 13.0-15.0   Rc 2450621  408
978  149.27889.7     18 21 23.04 -30 10 30.3 11.7-14.5   Rc 2450277  241
979  149.27889.43    18 21 29.41 -30 11 30.7 12.8-16.7   Rc 2450306  450
980  136.27916.215   18 21 30.59 -28 20 38.5 14.3-<20.4  Rc 2451357  305
981  136.28044.51    18 21 39.11 -28 30 10.4 14.0-<17.7  Rc 2451373  269
982  136.28047.41    18 21 51.87 -28 17 18.3 >12.0-<15.3 Rc 2450895  387
983  136.28171.57    18 22 12.08 -28 42 18.5 13.5-17.7   Rc 2450285  398
984  136.28430.126   18 22 36.46 -28 45 48.3 >11.5-16.4  Rc 2450585  305
985  153.28397.1410  18 22 44.41 -30 57 23.6 16.3-18.6   Rc 2451310  427
986  136.28561.2314  18 22 56.95 -28 41 17.6 15.6-<20.0  Rc 2451280: 398:
987  136.28566.126   18 23 03.24 -28 22 16.7 12.0-16.1   Rc 2451372  414
988  136.28820.47    18 23 36.01 -28 43 57.6 12.0-15.0   Rc 2450242  448
989  136.28824.483   18 23 36.96 -28 30 52.3 >14.0-<17.5 Rc 2451327  423
990  150.28922.21    18 23 51.64 -30 36 35.9 11.0-14.1   Rc 2450997  240
991  150.28927.26    18 23 53.52 -30 18 42.3 11.3:-15.3  Rc 2451318: 276:  Cl*NGC 6624 V1, b
992  150.28928.5     18 24 02.41 -30 13 33.6 11.3:-15.0  Rc 2449425: 335
993  150.29053.55    18 24 11.22 -30 34 35.0 11.9-15.7   Rc 2450948  320
994  150.29315.15    18 24 55.68 -30 25 44.5 11.7-15.5   Rc 2450582  292
995  166.30683.260   18 28 00.04 -25 52 22.4 13.9-17.5   Rc 2450640  463
996  147.31143.28    18 29 02.44 -29 53 16.3 12.2-<15.1  Rc 2450999  293
997  303.44071.35    18 29 28.84 -15 18 03.4 14.5-<19.0  Rc 2451329  383  TSVSC1 TN-S300112321-306-67-2
998  303.44578.145   18 30 15.88 -15 03 16.0 16.1-20.5:  Rc 2451397  512
999  303.44581.276   18 30 19.57 -14 52 46.2 >14.8-<17.5 Rc 2451054  422
1000 303.45088.65    18 30 57.37 -14 41 06.9 12.9-<15.5  Rc 2451320  375  TSVSC1 TN-S300112130-888-67-2
1001 303.45086.512   18 30 58.76 -14 50 32.8 14.0-18.1   Rc 2451366  412
1002 147.32054.11    18 31 07.93 -29 50 37.9 10.5:-14.6  Rc 2450999  292  GSC 06869-00888
1003 301.45108.136   18 31 08.40 -13 19 11.2 15.0-17.7   Rc 2451431  346
1004 301.45278.157   18 31 16.56 -13 11 59.2 15.5-<18.5  Rc 2450986  362
1005 302.45268.365   18 31 20.99 -13 54 39.1 14.8-18.0   Rc 2451237  311
1006 301.45269.57    18 31 23.06 -13 48 22.3 >14.3-<17.5 Rc 2450947  265
1007 302.45261.155   18 31 23.31 -14 19 08.7 16.0-19.8:  Rc 2451370  185
1008 302.45264.936   18 31 24.62 -14 08 01.9 14.3:-18.5  Rc 2451337  330
1009 301.45447.41    18 31 31.56 -13 10 33.4 >14.8-19.4: Rc 2449775: 363
1010 301.45447.87    18 31 41.60 -13 10 48.2 >14.9-<18.7 Rc 2451260  312
1011 301.45442.409   18 31 43.84 -13 30 44.5 14.2-17.1   Rc 2451073  257
1012 301.45611.486   18 31 50.52 -13 24 08.9 >13.5-18.0  Rc 2449977  376
1013 302.45600.25    18 31 57.94 -14 07 05.6 >13.9-<17.2 Rc 2451283  349
1014 302.45594.500   18 31 58.40 -14 33 26.6 14.0-17.9   Rc 2451350  349
1015 303.45586.33    18 32 00.92 -15 04 04.3 12.6-<17.5  Rc 2451401  370  TSVSC1 TN-S300112121-277-67-2
1016 301.45949.218   18 32 20.53 -13 18 12.9 13.9-17.5   Rc 2451340  254
1017 301.46108.40    18 32 51.42 -13 53 06.7 14.3-17.8:  Rc 2451044  388
1018 302.46270.66    18 32 57.06 -14 15 24.3 11.0:-15.0  Rc 2451374  276  TSVSC1 TN-S300113102-63-67-2
1019 301.46284.33    18 32 59.65 -13 21 38.4 14.0-17.7   Rc 2450320  409
1020 301.46276.189   18 32 59.97 -13 53 26.2 >13.8-16.3  Rc 2451180: 364
1021 301.46448.926   18 33 11.46 -13 38 25.0 14.4-17.8   Rc 2451418  344
1022 301.46451.1111  18 33 14.70 -13 23 30.6 15.3-18.7   Rc 2450545  484
1023 301.46453.72    18 33 16.15 -13 17 14.7 13.7-<19.0: Rc 2450547  311
1024 301.46445.287   18 33 26.36 -13 47 02.7 14.5-17.4   Rc 2450575  286
1025 301.46784.864   18 33 55.94 -13 36 09.1 14.2-18.0   Rc 2450601  507
```

Remarks:

a       Period based on a combination of MACHO Rc and ASAS-3 V data.

b       Contained in the globular cluster NGC 6624 and discovered as a variable object by Laborde and Fourcade (1966). It is identified here as a Mira variable for the first time.

c       Period based on a combination of MACHO Rc and OGLE Ic data.

**Acknowledgements:** This paper utilizes public domain data obtained by the MACHO Project, jointly funded by the US Department of Energy through the University of California, Lawrence Livermore National Laboratory under contract No. W-7405-Eng-48, by the National Science Foundation through the Center for Particle Astrophysics of the University of California under cooperative agreement AST-8809616, and by the Mount Stromlo and Siding Spring Observatory, part of the Australian National University. This research has made use of the SIMBAD and VizieR databases operated at the Centre de Données Astronomiques (Strasbourg) in France, of the Smithsonian / NASA Astrophysics Data System, of the AAVSO International Variable Star Index (VSX) and of the Two Micron All Sky Survey (2MASS). It is a pleasure to thank John Greaves and Patrick Wils for suggestions and helpful comments. We especially thank Sebastián Otero for his invaluable help in improving on the data of the MACHO sample.